\DocumentMetadata{pdfstandard=A-2b, lang=en-US}
\documentclass[aps,prd,notitlepage,nofootinbib,floatfix,letterpaper]{revtex4-2}

\usepackage[utf8]{inputenc}
\usepackage[T1]{fontenc}
\usepackage[american]{babel}
\usepackage{amsmath,amssymb}
\usepackage{graphicx}
\usepackage{booktabs}
\usepackage{siunitx}
\usepackage{xcolor}
\usepackage[colorlinks,linkcolor=blue,citecolor=blue,urlcolor=blue]{hyperref}

\newcommand{\beamEnergy}{924.2}
\newcommand{\laserWavelength}{1030}
\newcommand{\gammaZero}{1808.6}

\newcommand{\aZero}{0.5}
\newcommand{\lFill}{4.39}
\newcommand{\coElectronsShot}{1e+09}
\newcommand{\coChargePc}{160}
\newcommand{\coRepHz}{1000}

\newcommand{\coConeShot}{3.97e+09}
\newcommand{\svRepHz}{100000}

\newcommand{\repHzThirteen}{10000}
\newcommand{\dwellCap}{347}

\newcommand{\surveyRange}{2.911}
\newcommand{\surveyVrel}{9.433}
\newcommand{\repHzFourteen}{100000}
\newcommand{\drFireMeanT}{0.0079}
\newcommand{\drFirePerDayT}{3.43e-03}
\newcommand{\svFireMeanT}{0.211}
\newcommand{\svFirePerDayT}{0.220}
\newcommand{\discLkm}{0.2}
\newcommand{\discElectrons}{1e+09}

\newcommand{\sigLkm}{0.2}
\newcommand{\sigElectrons}{1e+09}
\newcommand{\sigAdet}{1}

\newcommand{\sigHardTot}{0.117}
\newcommand{\sigSoftTot}{0.217}
\newcommand{\sigCapTot}{0.230}
\newcommand{\sigGamTot}{0.0297}
\newcommand{\sigHardLate}{1.9e-05}
\newcommand{\sigSoftLate}{6.1e-05}
\newcommand{\sigCapLate}{0.071}
\newcommand{\sigGamLate}{0.0043}
\newcommand{\sigEnvHard}{0.112}
\newcommand{\sigEnvSoft}{0.142}
\newcommand{\sigEnvCap}{0.134}
\newcommand{\sigEnvGam}{2.34}
\newcommand{\nSteady}{7,852}

\newcommand{\nUnshelled}{37}

\newcommand{\raanBinDeg}{10}
\newcommand{\sThreeInc}{53.16}
\newcommand{\sThreeAltA}{358}
\newcommand{\sThreeAltB}{486}
\newcommand{\sThreeAltC}{552}
\newcommand{\sThreeAltD}{573}
\newcommand{\sThreeRaanA}{0}
\newcommand{\sThreeRaanB}{90}
\newcommand{\sThreeRaanC}{180}
\newcommand{\sThreeRaanD}{270}
\newcommand{\sTwoInc}{53.16}
\newcommand{\sTwoRaan}{0}
\newcommand{\sTwoHops}{38}
\newcommand{\sTwoAltLo}{340}
\newcommand{\sTwoAltHi}{575}
\newcommand{\lambdaStar}{9.56}

\newcommand{\reachCoStatic}{1.76}
\newcommand{\gammaReachCoStatic}{0.55}
\newcommand{\repRangeThirteen}{0.71}
\newcommand{\fullServeTen}{115}
\newcommand{\fullServeFifty}{673}
\newcommand{\fullServeNinety}{1538}

\newcommand{\driftServeTen}{59}
\newcommand{\driftServeFifty}{343}
\newcommand{\driftServeNinety}{870}
\newcommand{\surveyServeTen}{179}
\newcommand{\surveyServeFifty}{884}
\newcommand{\surveyServeNinety}{1630}

\newcommand{\svOverCap}{21}
\newcommand{\svDuty}{2.5e-06}
\newcommand{\tenthTargets}{785}

\newcommand{\fullFleet}{69}

\newcommand{\gammaDoseClose}{0.61}
\newcommand{\gammaReachClose}{0.7}
\newcommand{\reconLaunches}{48}
\newcommand{\orbitalLaunches}{121}

\newcommand{\reconPercentNow}{5}
\newcommand{\sigmaTheta}{0.30}
\newcommand{\energySpread}{0.03}
\newcommand{\nScatterMean}{10.9}
\newcommand{\yieldGenerated}{10.903}
\newcommand{\yieldCone}{3.967}
\newcommand{\coneFraction}{0.364}
\newcommand{\gammaLastOverZero}{0.928}
\newcommand{\rWOneCasing}{21.00}
\newcommand{\mWOneCasing}{14.3}
\newcommand{\rWOneHe}{20.00}
\newcommand{\mWOneHe}{52.8}
\newcommand{\rWOneTamper}{10.00}
\newcommand{\mWOneTamper}{52.3}
\newcommand{\rWOneRefl}{7.00}
\newcommand{\mWOneRefl}{1.7}
\newcommand{\rWOneCore}{5.00}
\newcommand{\mWOneCore}{4.0}
\newcommand{\rWOneCavity}{4.25}
\newcommand{\mWOneCavity}{3.87e-04}
\newcommand{\rWTwoCase}{21.50}
\newcommand{\mWTwoCase}{31.0}
\newcommand{\rWTwoGap}{20.50}
\newcommand{\mWTwoGap}{6.55e-03}
\newcommand{\rWTwoLiner}{20.00}
\newcommand{\mWTwoLiner}{20.2}
\newcommand{\rWTwoCav}{19.90}
\newcommand{\mWTwoCav}{3.04e-02}
\newcommand{\rWTwoPriHe}{19.44}
\newcommand{\mWTwoPriHe}{52.8}
\newcommand{\rWTwoPriRefl}{7.00}
\newcommand{\mWTwoPriRefl}{1.7}
\newcommand{\rWTwoPriCore}{5.00}
\newcommand{\mWTwoPriCore}{4.0}
\newcommand{\rWTwoPriCavity}{4.25}
\newcommand{\mWTwoPriCavity}{3.87e-04}
\newcommand{\rWTwoSecFoam}{19.90}
\newcommand{\mWTwoSecFoam}{0.4}
\newcommand{\rWTwoSecPusher}{15.00}
\newcommand{\mWTwoSecPusher}{16.8}
\newcommand{\rWTwoSecFuel}{14.68}
\newcommand{\mWTwoSecFuel}{10.9}
\newcommand{\rWTwoSecPlug}{3.35}
\newcommand{\mWTwoSecPlug}{3.0}
\newcommand{\rDuBall}{9.10}
\newcommand{\mDuBall}{60.0}

\newcommand{\isoPuNine}{95.19}
\newcommand{\isoPuZero}{4.71}
\newcommand{\isoPuOne}{0.10}
\newcommand{\isoAm}{0.2}
\newcommand{\isoNatU}{0.7}
\newcommand{\isoHEU}{95}
\newcommand{\isoDU}{0.2}
\newcommand{\isoLiSix}{95}
\newcommand{\rhoPu}{19.8}
\newcommand{\rhoNatU}{19.0}
\newcommand{\rhoHEU}{19.0}
\newcommand{\rhoDU}{19.0}
\newcommand{\rhoTATB}{1.80}
\newcommand{\rhoLiD}{0.83}
\newcommand{\rhoFoam}{0.022}
\newcommand{\rhoAl}{2.699}
\newcommand{\rhoBe}{1.848}
\newcommand{\rhoAir}{1.20e-03}
\newcommand{\busXBus}{410}
\newcommand{\busYBus}{260}
\newcommand{\busZBus}{20}
\newcommand{\busRhoBus}{0.195}
\newcommand{\busNBus}{1}
\newcommand{\busMassBus}{360}
\newcommand{\busFBusAl}{1.00}
\newcommand{\busXAvionics}{120}
\newcommand{\busYAvionics}{80}
\newcommand{\busZAvionics}{16}
\newcommand{\busRhoAvionics}{0.977}
\newcommand{\busNAvionics}{1}
\newcommand{\busMassAvionics}{150}
\newcommand{\busFAvionicsCu}{0.35}
\newcommand{\busFAvionicsC}{0.25}
\newcommand{\busFAvionicsO}{0.20}
\newcommand{\busFAvionicsAl}{0.10}
\newcommand{\busFAvionicsSi}{0.10}
\newcommand{\busXBattery}{100}
\newcommand{\busYBattery}{60}
\newcommand{\busZBattery}{14}
\newcommand{\busRhoBattery}{0.952}
\newcommand{\busNBattery}{1}
\newcommand{\busMassBattery}{80}
\newcommand{\busFBatteryO}{0.25}
\newcommand{\busFBatteryC}{0.20}
\newcommand{\busFBatteryAl}{0.15}
\newcommand{\busFBatteryCu}{0.15}
\newcommand{\busFBatteryFe}{0.10}
\newcommand{\busFBatteryF}{0.05}
\newcommand{\busFBatteryLi}{0.05}
\newcommand{\busFBatteryP}{0.05}
\newcommand{\busXArgon}{60}
\newcommand{\busYArgon}{60}
\newcommand{\busZArgon}{14}
\newcommand{\busRhoArgon}{0.794}
\newcommand{\busNArgon}{1}
\newcommand{\busMassArgon}{40}
\newcommand{\busFArgonAr}{1.00}
\newcommand{\busXAntenna}{400}
\newcommand{\busYAntenna}{250}
\newcommand{\busZAntenna}{5}
\newcommand{\busRhoAntenna}{0.080}
\newcommand{\busNAntenna}{1}
\newcommand{\busMassAntenna}{40}
\newcommand{\busFAntennaCu}{0.31}
\newcommand{\busFAntennaC}{0.25}
\newcommand{\busFAntennaO}{0.25}
\newcommand{\busFAntennaSi}{0.15}
\newcommand{\busFAntennaH}{0.04}
\newcommand{\busXWing}{1250}
\newcommand{\busYWing}{420}
\newcommand{\busZWing}{2}
\newcommand{\busRhoWing}{0.033}
\newcommand{\busNWing}{2}
\newcommand{\busMassWing}{70}
\newcommand{\busFWingO}{0.35}
\newcommand{\busFWingSi}{0.30}
\newcommand{\busFWingAl}{0.15}
\newcommand{\busFWingC}{0.15}
\newcommand{\busFWingH}{0.05}

\newcommand{\repCoorbit}{1000}
\newcommand{\panelX}{410}
\newcommand{\panelY}{260}
\newcommand{\solarPhiMV}{650}

\newcommand{\voxelCm}{1}
\newcommand{\liFraction}{0.01}

\newcommand{\faceSurvey}{3.16}
\newcommand{\blankShot}{2}
\newcommand{\blankReturn}{1}
\newcommand{\doseCap}{0.25}
\newcommand{\firingHalfAngle}{45}
\newcommand{\lMaxKm}{10}
\newcommand{\faceCoorbit}{1}
\newcommand{\dvCoorbit}{5}
\newcommand{\dvSurvey}{0}

\newcommand{\gammaWinLo}{0.1}
\newcommand{\gammaWinHi}{12}
\newcommand{\repSurvey}{100000}

\newcommand{\driveJ}{4}
\newcommand{\scatterJ}{4}
\newcommand{\etaElectric}{0.5}
\newcommand{\electricJ}{\fpeval{(\driveJ+\scatterJ)/\etaElectric}}
\newcommand{\bb}{\bar{\beta}}

\begin{document}

\title{Verification of the Outer Space Treaty with active interrogation}

\author{Patrick Huber}
\email{pahuber@vt.edu}
\affiliation{Center for Neutrino Physics, Department of Physics, Virginia Tech, Blacksburg, VA 24061, USA}

\date{\today}

\begin{abstract}
We present an analysis of active interrogation in space with the goal
to verify the absence of nuclear weapons hidden in a large
constellation of satellites. We propose a 16\,MeV photon beam
generated by advanced laser wakefield accelerators and the detection
of fission neutrons using segmented organic scintillators. We use
Starlink as a real-world example for the constellation and solve the
inspector routing problem against this set of 7,852 satellites over a
5-year mission period. We perform detailed simulation of the nuclear
and orbit physics: signal generation and detection with GEANT4 against
an orbit-resolved neutron and gamma background, routes flown with
Orekit including the effects from the Earth's oblateness, and detection
decided by a likelihood ratio test. We identify different encounter
types and fleet designs based on a range of inspector platforms. The
radiation dose to target satellites from the beam can be kept to about
1\% or less of the 5-year space environment dose. We find that a small number
of inspection platforms, 20--70 depending on the accelerator
assumptions, can serve inspections to the full constellation with a
detection probability of 90\% against a single deployed warhead. A
10\% random sample of the whole fleet can be inspected by 2--20
platforms providing a 37\% detection probability against 5 deployed
warheads.

\end{abstract}

\maketitle

\section{Introduction}
\label{sec:intro}

On July 9, 1962, during the Starfish Prime test a 1.4\,Mt thermonuclear
device was detonated at an altitude of 400\,km and led to the formation of
a belt of trapped electrons~\cite{Stassinopoulos:2015,Hess:1963}. As a
consequence, at least eight of the 24 satellites then in orbit
or launched in the weeks that followed sustained
damage of varying degrees~\cite{Conrad:2010}, demonstrating
the effectiveness of nuclear detonations as an anti-satellite weapon.
This was followed in 1967 by the adoption of the Outer Space Treaty
(OST), whose article IV commits States Parties ``not to place in orbit
around the Earth any objects carrying nuclear weapons or any other
kinds of weapons of mass destruction''~\cite{ost1967}. Among the reasons the OST has no
verification clause are that at that time no technology for in-space
verification was apparent and on-the-ground inspection was seen as
problematic vis-\`a-vis classified reconnaissance satellite
programs~\cite{Porteous:2025}.

The space domain has changed considerably in the intervening nearly
six decades; there are now 16,946 payloads among the 44,964 cataloged
objects in orbit~\cite{ESA:2026}, serving a wide range of missions,
most of them not related at all to military purposes. In 2024 the
world downstream market --- satellite communications, navigation and
Earth observation --- was worth \texteuro408 billion (about \$465
billion\footnote{at the September 2026 exchange rate.})~\cite{ESA:2025econ}. Space also remains strategically important for
essentially the same class of applications that generate economic
value: communication, navigation and Earth observation. Nuclear command,
control and communications depends on the same domain: missile warning and attack
characterization come from the Space-Based Infrared System in geosynchronous and highly
elliptical orbits, now moving to a proliferated architecture in low and medium Earth orbit, and
assured communication to nuclear forces from the hardened Advanced Extremely High Frequency
satellites~\cite{CRS:NC3}.

There has been a keen interest by the U.S., Russia, China and India in
developing kinetic anti-satellite weapons for a while now; they have each
destroyed a satellite in orbit with a direct-ascent weapon: the
United States in 1985 and again in 2008, China in 2007, India in 2019,
and Russia on November 15, 2021, against its own
Kosmos~1408~\cite{Kimball:2022,Bugos:2021b}. Much of the growth in low
Earth orbit is in proliferated constellations, where relatively
low-cost, mass-produced satellites provide a highly resilient
architecture with global coverage at all times. The failure of a
single satellite, or of a fraction of them, leaves the overall
capability intact, a key consideration for {\it e.g.} internet
connectivity. This also makes these constellations virtually immune to
direct-ascent kinetic attacks. That property is the reason the United
States adopted proliferated architectures for resilience, and the
reason Chinese analysts see them as removing the key nodes their own
operational concept depends on~\cite{RAND:2025}. These constellations do remain
vulnerable to nuclear attack: a single detonation reaches about 20\%
of the low Earth orbit band promptly, and delayed effects from trapped
electrons can damage satellites for months to
years~\cite{Snyder:2025}. However, the effects of the trapped radiation are
difficult to predict, as experienced during Starfish
Prime~\cite{Stassinopoulos:2015}. Historically, U.S. damage
calculations have counted the effects that could be modeled with
confidence and left the others out altogether~\cite{Eden:2005}. 
Stationing nuclear weapons in space for anti-satellite purposes,
instead of relying on direct ascent for delivery, could be attractive
if one is faced with effective, potentially space-based, anti-missile
defenses~\cite{CSR:2026}. Given these developments,
verification of the OST seems more needed than ever. Recently, Kosmos~2553 has attracted a lot
of speculation about its role in the development of a nuclear space
weapon capability by
Russia~\cite{Stewart:2024,Sullivan:2024,SWF:ASAT}. More significantly,
Russia vetoed a U.N. Security Council resolution that reaffirmed
that parties to the OST shall not place nuclear weapons in orbit, and
China abstained~\cite{Kuramitsu:2024}.

Starlink, operated by SpaceX, is currently the largest constellation
with the catalog of March 25, 2026, holding
\nSteady{} of its satellites in steady operation.  We will use it as
the reference target set throughout this work, see
Sec.~\ref{sec:starlink}. The war in Ukraine has clearly blurred the
distinction between civilian and military space systems: a commercial
constellation donated for civilian connectivity became a battlefield
communications link, its operator set limits on offensive use on its
own authority, and the Pentagon then bought the same service under a
contract~\cite{Abels:2024}. Russia told the First Committee of the
U.N.  General Assembly that ``quasi-civilian
infrastructure may become a legitimate target for
retaliation''~\cite{Vorontsov:2022} and Chinese military analysts
assess Starlink as a military system and call for a constellation of
their own~\cite{RAND:2025}. Thus, any future verification of the OST
will need to work in the context of large commercial constellations,
and focusing on single platforms, orbits or designated satellites will
fall short of the task --- if not in the U.S. perception, then in the
Chinese or Russian assessment.

The concern we address is therefore a warhead placed in orbit among
the satellites of a large constellation, in or disguised as one of
them, and the task of verification is to give the other parties
confidence that none of the satellites carries one. Because active
interrogation delivers a radiation dose and requires coordinated
close approaches of sufficient dwell time, we assume a cooperative
framework under a negotiated agreement.

In summary, space is an ever more important domain, both economically and
strategically, and an ever more contested one. One or a few nuclear explosions in space
today would have wide-ranging and difficult-to-predict consequences in
terms of economic, political, tactical and strategic impact; these
consequences are likely becoming more severe in the future. At the
same time the pressure on the OST seems to be increasing, and the line
between military and peaceful uses, and between commercial and national assets,
is becoming less well defined. Thus, the question of verification is
gaining renewed interest. There have been two recent studies, one by
Porteous~\cite{Porteous:2025} where a broad overview of the technical
feasibility of in-space verification was presented and the other by
Danagoulian~\cite{danagoulian} that contains a specific proposal to
address the purpose of Kosmos~2553 by exploiting its unusual,
high-radiation orbit. Both have in common that they focus on one or
a few target satellites. Here we would like to expand the subject to a
large set of satellites, exemplified by Starlink, and specifically
understand whether active interrogation can be useful in this context.

We simulate signal generation and detection with GEANT4~\cite{Agostinelli:2003}
against an orbit-resolved neutron and gamma background, plan inspection routes
through the Starlink catalog and fly them with Orekit~\cite{orekit} including
the effects of the Earth's oblateness, and decide detection with a likelihood ratio test.  Inspecting all 7,852
Starlink satellites within five years at a 90\% detection probability takes
69 co-orbiting inspectors (S1), or four passive fly-by platforms (S3)
supported by 16 inspectors.  Inspecting a random tenth of the constellation
takes 17 inspectors, or a single maneuvering fly-by platform (S2) and one
inspector, and gives a party that hides five warheads a 37\% risk of
discovery.  No inspected satellite receives more than 0.25\,Gy, less than
1\% of what it accumulates from the natural radiation environment over five
years.  Finally, we identify the minimum set of necessary treaty deliverables.

\section{Active interrogation in space}

Active interrogation on Earth is a well-studied
subject~\cite{Fetter:1990,Runkle:2012,Geddes:2015} and many
experimental demonstrations have been conducted. The basic idea is
simple: use an external beam of radiation to induce a specific
radiation response from the object under test. By carefully choosing the type of
interrogating and emitted radiation, high
specificity to various attributes can be achieved, {\it e.g.} high
atomic number, high density, fissile material and even isotopic
composition~\cite{Bertozzi:2005,Vavrek:2018}. Each leg of the
radiation propagation in general will suffer its own $1/r^2$ geometric
acceptance penalty; the overall scaling of the detected signal is
unfavorable, $1/r^4$, and the simple expedient of increasing the strength
of the interrogating beam will eventually run into radiation
protection issues for either people or electronics. For short standoff
distances the interference of the operation of the radiation
source and/or the interrogating beam with the detection of emitted
radiation puts further constraints on the systems and tends to favor
signatures recorded after the interrogating pulse. Highly capable
systems are possible but tend to be complex and, for cargo and
standoff screening, have not gone beyond
demonstration~\cite{Hall:2007,Favalli:2025}; active neutron
interrogation is routinely fielded in waste assay~\cite{ASTM:C1493},
and only radiography-type systems and x-ray backscatter have found
widespread use in cargo and vehicle inspection~\cite{Geddes:2017}.

The space environment offers some advantages for active interrogation.
It is a high-radiation environment without humans present, and
thus dose caps can be higher than on Earth, set by the tolerance of
moderately radiation-hard electronics, see Sec.~\ref{sec:signatures}. The space vacuum allows neutral
beams to propagate long distances without scattering or absorption;
thus, well-collimated beams can help to improve the overall scaling
from $1/r^4$ to $1/r^2$, see Sec.~\ref{sec:source}. The resulting longer time of flight combined
with the absence of scattering separates the beam radiation from the
secondary radiation, so that, for instance, prompt neutrons can be
easily detected, see Sec.~\ref{sec:detector}.

Space-based active interrogation has also been studied in the context
of the Strategic Defense Initiative (SDI), where the goal was to identify
incoming nuclear warheads to separate them from decoys. The idea was to
accelerate protons or deuterium ions to a few hundred MeV and then to
neutralize them so that they can propagate as an extremely well-collimated
beam through the Earth's magnetic field~\cite{OTA:1988,APS:1987}. Once
intercepted by the target, they would create a spray of secondary
particles depending on the areal mass density of the target. The
delivered doses would have been in the hundreds of grays and the satellites
providing the beam would have been immense in size and power
consumption, although these systems might have been able to interrogate
targets at very large distances, 1,000\,km, in a very short time, tens
of milliseconds~\cite{OTA:1988,APS:1987}. Of course, the very same
beam could disable a target within seconds; for SDI this was considered
useful, or at least not detrimental.

In a treaty verification context, any interrogation beam source has
to be so weak as to pose no threat to the target even if fired at
short range over a long period of time at a target that has not been
extensively radiation hardened. We will consider a novel beam source,
inverse Compton scattering from a $\sim 1\,$GeV electron beam that is
produced by a laser wakefield accelerator
(LWFA)~\cite{Leemans:2006,Rykovanov:2014}. This results in a
16\,MeV photon beam that can effectively induce photofission and at
the same time delivers only a minimal radiation dose. Diode-pumped
fiber lasers can provide very good wall-plug electric efficiency~\cite{Kiani:2022} and the
average electric power of the system we present here is measured in
watts. Neutron detectors based on recent progress in neutrino
detection using segmented organic
scintillators~\cite{LiquidO:2019mxd,Haghighat:2018mve,MobileAntineutrinoDemonstratorProject:2025sna}
ensure efficient detection and at the same time reject charged-particle cosmic-ray
backgrounds. Given that treaty verification is a task that plays out
over years and not seconds, we can use maneuverable platforms to reduce
the standoff to a point where these low-power beam sources can be
effective.

\section{Photon source}
\label{sec:source}


One of the major drawbacks for active interrogation is that both the
interrogating beam and the signal each carry a $1/r^2$ penalty, leading
overall to a $1/r^4$ scaling that usually severely limits the useful
range~\cite{Fetter:1990}. One way to overcome this is to use a well-collimated beam:
as long as the beam under-fills the target, the flux that can be delivered to the target is
\emph{independent} of distance.  For a target of size $s$ and a beam divergence $\theta$, the
beam-fill distance is $L_{\rm fill} = s/\theta$.  This simple geometric fact puts a premium on a
well-collimated photon beam with a small divergence $\theta$.

The traditional way to produce photon beams at the 10\,MeV scale is bremsstrahlung from a
relativistic electron beam of energy $E_e$, which produces a secondary photon spectrum roughly of the form
$dN/d\eta \sim E_e/\eta$ with $\eta \leq 1$.  The mean photon energy is therefore within a
factor of order unity of the electron energy $E_e$.  At the same time, the opening angle of the
resulting beam is $1/\gamma_e$, where $\gamma_e$ is the Lorentz factor of the electron beam, and
is thus tens of milliradians.

Inverse Compton scattering (ICS), on the other hand, is a process in which an optical photon is
up-scattered in energy by colliding with a beam of relativistic electrons, resulting in the photon
spectrum
\begin{equation}
  E(\psi;\gamma) = \frac{4\gamma^2\hbar\omega_L}{1 + a_0^2/2 + \gamma^2\psi^2},
  \label{eq:ics}
\end{equation}
where $\omega_L$ is the frequency of the optical photons, $\psi$ is the angle relative to the
electron beam axis in the laboratory frame, and $a_0 = \aZero$ is the dimensionless strength of
the scattering laser, which accounts for nonlinear effects~\cite{Rykovanov:2014}.  We use an electron beam of $E_e = \SI{\beamEnergy}{\mega\electronvolt}$
($\gamma_0 = \gammaZero$),  scattered off a counter-propagating
\SI{\laserWavelength}{\nano\metre} Yb-fiber pulse ($\hbar\omega_L = \SI{1.2037}{\electronvolt}$).
The endpoint of the resulting spectrum, $4\gamma_0^2\hbar\omega_L$, is
\SI{\fpeval{round(4*\gammaZero^2*1.2037e-6,2)}}{\mega\electronvolt}.

With these choices the cone
half-angle is $1/\gamma_0 =
\SI{\fpeval{round(1000/\gammaZero,2)}}{\milli\radian}$.  For our case,
the beam just fills the bus panel of a Starlink V2-mini, of size
$\panelX\times\SI{\panelY}{\centi\metre}$, at $L_{\rm fill} =
\SI{\lFill}{\kilo\metre}$; beyond $L_{\rm fill}$ the intercepted
fraction falls as $(L_{\rm fill}/L)^2$.
The Monte Carlo treatment of the source, with multiple scattering,
electron recoil, pointing and energy spread, is given in App.~\ref{app:source}.

\begin{figure*}
\includegraphics[width=\textwidth]{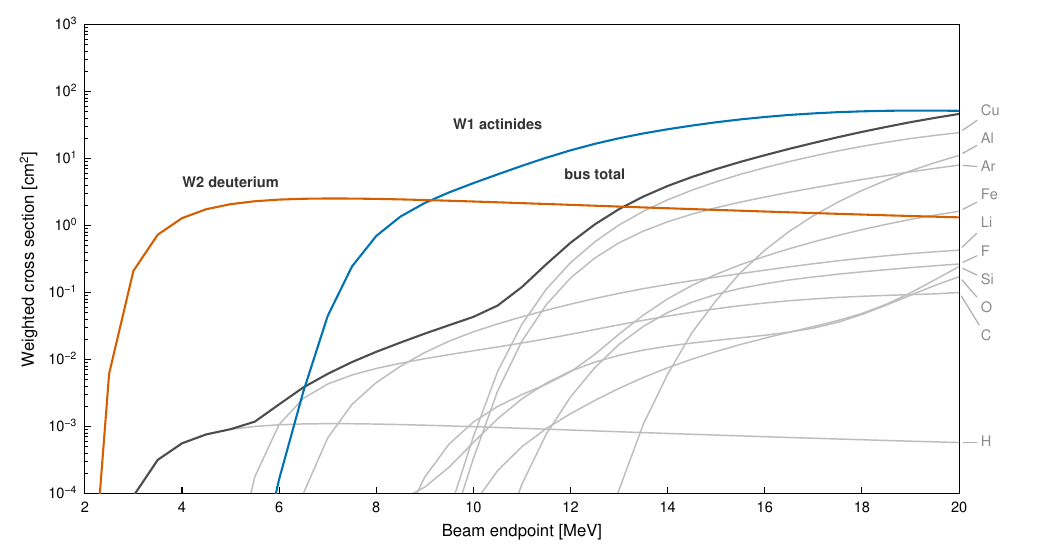}
\caption{Neutron production by element against the beam endpoint: atoms times the
neutron-yield cross section $\sigma_{1n} + 2\sigma_{2n} + 3\sigma_{3n}$, folded over the ICS
spectrum, with the endpoint shown on the x-axis.  Bus elements from the satellite inventory; W1 actinides; the
deuterium of W2's LiD secondary, for details see App.~\ref{app:models}.}
\label{fig:massweighted}
\end{figure*}

The photofission cross section of the relevant actinides peaks around
13--15\,MeV, at the giant dipole resonance~\cite{Caldwell:1980}.
However, a number of photo-induced processes in non-actinides also
produce neutrons, and it is therefore {\it a priori} not obvious which
beam energy provides the best discrimination.  The photoneutron
threshold of deuterium is as low as \SI{2.2}{\mega\electronvolt}, and
for the main structural materials these processes open between
\SI{5}{\mega\electronvolt} (C, O) and \SI{12.5}{\mega\electronvolt}
(Al), whereas the onset of actinide photofission is around
6\,MeV~\cite{iaea_pnd}.  In Fig.~\ref{fig:massweighted} we show the
neutron production cross section for each element on board the
satellite, weighted by its mass.  We account for each isotope
separately, with natural isotopic abundance for the structural
materials and the appropriate isotopic composition for the warheads,
as described in Sec.~\ref{sec:signatures}.  The deuterium mass
corresponds to that of the secondary stage of a thermonuclear device.
The quantity shown is the neutron-yield cross section $\sigma_{1n} +
2\sigma_{2n} + 3\sigma_{3n}$, with its full energy dependence, folded
with the ICS spectrum as explained in
  App.~\ref{app:source}, and multiplied by the number of atoms
present; the result is shown as a function of the endpoint
$4\gamma_0^2\hbar\omega_L$.  Below 6\,MeV only the deuterium responds;
between 6 and about \SI{14}{\mega\electronvolt} the actinide curve
rises fastest and stands an order of magnitude above the bus; above
the resonance the actinide yield saturates while the bus keeps rising
as the heavier structural elements open.  Our endpoint of
\SI{\fpeval{round(4*\gammaZero^2*1.2037e-6,2)}}{\mega\electronvolt}
was chosen using the actual likelihood and detector performance and is
driven by the environmental backgrounds.

We assume an electron beam energy of \SI{\beamEnergy}{\mega\electronvolt} and
\num[print-unity-mantissa=false]{\coElectronsShot} electrons per shot, corresponding to \SI{\coChargePc}{\pico\coulomb},
which yields \num{\coConeShot} photons per shot inside the cone $\gamma_0\theta_{\rm lab} \le 1$.
The same charge per shot serves every platform; the platforms differ only in repetition rate,
\SI{\fpeval{\coRepHz/1000}}{\kilo\hertz} for S1, \SI{\fpeval{\repHzThirteen/1000}}{\kilo\hertz} for S2
and \SI{\fpeval{\svRepHz/1000}}{\kilo\hertz} for S3. We will define these platforms in Tab.~\ref{tab:platforms}.

Single-stage beams near \SI{1}{\giga\electronvolt} that deliver
\SI{30}{\pico\coulomb} have been demonstrated~\cite{Leemans:2006};
more recently, \SIrange{780}{840}{\mega\electronvolt} at
\SIrange{8.5}{23.6}{\pico\coulomb} with \SIrange{0.24}{0.41}{\percent}
rms energy spread and a beam divergence of
\SIrange{0.1}{0.4}{\milli\radian} have been achieved~\cite{Ke:2021}.
With close to optimal beam loading, higher charges of \SIrange{220}{300}{\pico\coulomb}
have been reached at
\SI{250}{\mega\electronvolt}~\cite{Couperus:2017}.  Thus our
parameters in terms of charge and beam energy are within the
demonstrated realm individually and moderately optimistic in
combination. Scaling of the repetition rate into the \SIrange{1}{100}{\kilo\hertz} range is needed
for applications in high-energy physics as well, and the kBELLA
project~\cite{Kiani:2022} aims to demonstrate this with diode-pumped
solid-state fiber lasers, which will also improve the wall-plug
electric efficiency.

Each shot needs two laser pulses, a drive pulse that makes the wake and a scattering pulse that provides the photons for the inverse Compton scattering.
The drive pulse energy of \SI{\driveJ}{\joule} is derived in App.~\ref{app:source}.

The scattering pulse intensity at the focus determines how many high-energy photons per electron can be achieved. We take
a scattering pulse of \SI{\scatterJ}{\joule}, scaled from Rykovanov's photofission design
point~\cite{Rykovanov:2014}; that is \SI{\fpeval{\driveJ+\scatterJ}}{\joule}
of light per shot.
With an electric-to-optical efficiency of \SI{\fpeval{round(100*\etaElectric,0)}}{\percent}, the
upper end of what fiber lasers reach~\cite{Kiani:2022}, this corresponds to \SI{\electricJ}{\joule} of
electricity per shot, shared equally between the drive and the scattering pulse.

\newcommand{\coDaysPerTarget}{\fpeval{1808.5/(785/17)}}
Electric power while firing is energy per shot times the repetition
rate. The total number of shots aimed at a target is capped either by
the total radiation dose or by the time the target is in range. For the three platforms S1, S2 and S3 that will be introduced in Sec.~\ref{sec:orbits}, the
resulting peak power ranges from \fpeval{round(\electricJ*\repCoorbit/1e3,0)}, \fpeval{round(\electricJ*\repHzThirteen/1e3,0)}, and \num[group-minimum-digits=4,group-separator={,}]{\fpeval{round(\electricJ*\repHzFourteen/1e3,0)}}\,kW,
respectively. While these are large numbers, each encounter is short,
and the average energy per target is \num[group-minimum-digits=4,group-separator={,}]{\fpeval{round(\dwellCap*\repCoorbit*\electricJ/1e3,0)}}, \fpeval{round(\electricJ*\repHzThirteen*\drFireMeanT/1e3,1)} and \fpeval{round(\electricJ*\repHzFourteen*\svFireMeanT/1e3,0)}\,kJ, since the
dwell times are very different for the three platforms. This is the
energy a battery would have to provide over the encounter. The average power is then
determined by the duty factor and we obtain \fpeval{round(\dwellCap*\repCoorbit*\electricJ/(\coDaysPerTarget*86400),1)}, \fpeval{round(\electricJ*\repHzThirteen*\drFirePerDayT/86400,3)} and \fpeval{round(\electricJ*\repHzFourteen*\svFirePerDayT/86400,1)}\,W.

For scale, the battery pack of a Tesla
Model~S Plaid holds \SI{100}{\kilo\watt\hour}
(\SI{3.6e5}{\kilo\joule}) with a mass of \SI{479}{\kilo\gram} and
feeds a \SI{760}{\kilo\watt} drivetrain~\cite{TeslaPlaid:2026}; these
energy densities are matched by flight-qualified lithium-ion
cells~\cite{NASA:SmallSatPower}. This indicates that commercially available
batteries can provide both sufficient stored energy and sufficient
peak power for the needs of this mission. For S3, each
firing event lasts \SIrange{0.1}{1}{\second}, and this may be served
more efficiently by ultracapacitors~\cite{Skeleton:2026}.  We also
note that the average power stays small, at most a few Watts in all
cases, so neither energy generation by a solar array nor heat
rejection should present major design issues. This analysis also
indicates that variations of our assumptions about the efficiency do not
present a major risk for the feasibility. The one key ingredient
needed is a flight-qualified multi-joule, \SIrange{10}{50}{\femto\second} fiber laser system
that can achieve repetition rates of \SIrange{1}{10}{\kilo\hertz}, maybe even \SI{100}{\kilo\hertz}.

\section{Signatures \& backgrounds}
\label{sec:signatures}

\begin{figure}
\includegraphics[width=0.75\textwidth]{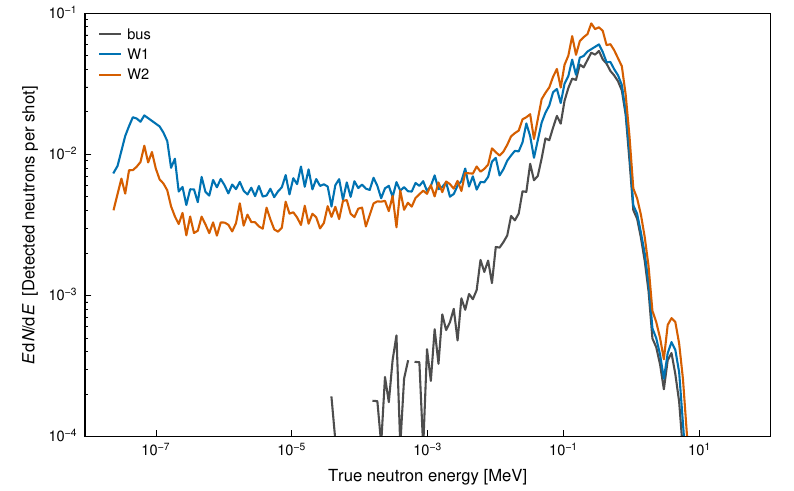}
\caption{Detected neutrons weighted by energy against the  neutron energy for one shot of \num[print-unity-mantissa=false]{\discElectrons} electrons at
\SI{\discLkm}{\kilo\metre}, for the bare bus and for W1 and W2 inside the bus. }
\label{fig:discriminant}
\end{figure}

For the discussion here we will consider an empty Starlink bus
(the V2 Mini, Ref.~\cite{spacex_fcc_2022_exhibitB}; size and materials in
Tab.~\ref{tab:bus}), labeled ``bus'', and a
one-stage fission device labeled ``W1'', and a two-stage thermonuclear
device labeled ``W2'', both with design specifications drawn from
Ref.~\cite{Sonder:2023} and shown in Fig.~\ref{fig:objects}. The detailed
isotopic compositions can be found in Tabs.~\ref{tab:objects} and \ref{tab:materials} in
App.~\ref{app:models}. The
15.75\,MeV photon beam will induce photo-fission and
$(\gamma,n)$-reactions both leading to the production of neutrons.
The simulation of the neutron and gamma emission and the bus model are
described in App.~\ref{app:models}. In
Fig.~\ref{fig:discriminant} we show the energy distribution of the
resulting neutron signals, and can clearly observe that fast neutron
yields between W1/W2 and the bus are fairly similar, albeit on a
logarithmic scale. However, epithermal and slower neutrons are absent
from the bus-only case, but are copious for W1 and W2. This highlights
an important fact: we are not looking at bare masses of fissile
materials; instead, we see a moderated fission neutron source where the
moderation is provided by the shell of high-explosive surrounding the
fissile core. We find that the $k_{\rm eff}$ of W1 and W2 is in the range of 0.62--0.67 corresponding to a neutron multiplication of 2.7--3.0.
The moderation effect is somewhat less pronounced for W2 since the
radiation case is indeed bare and we see its neutrons
unmoderated. Note that, for instance, the presence of large amounts of
fuel containing hydrogen would not lead to the same level of
moderation, since the fuel would sit next to the fissile material and
would not surround it in a $4\pi$ geometry.

\begin{figure}
\includegraphics[width=0.8\textwidth]{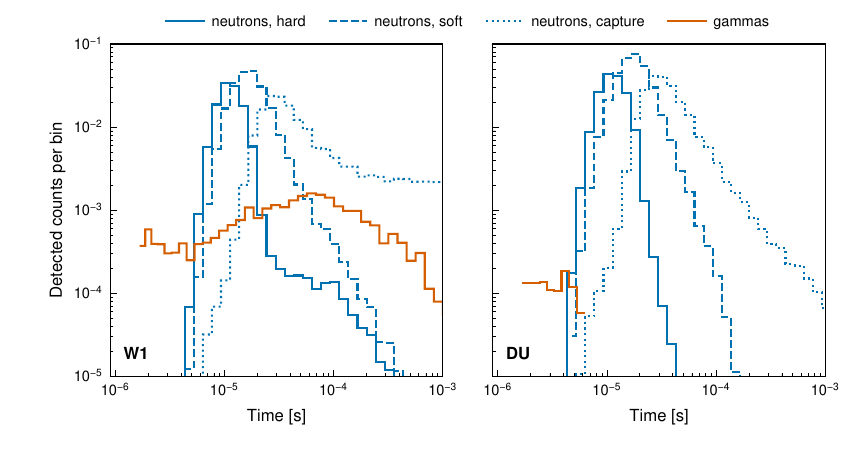}
\caption{Detected counts per arrival-time bin from one pulse of \num[print-unity-mantissa=false]{\sigElectrons} electrons at
\SI{\sigLkm}{\kilo\metre} on W1 and on the DU ballast, on the \SI{\sigAdet}{\metre\squared}
S1 detector. Shown are the three neutron
classes and the gamma channel.  Time zero is defined as the beam reaching the target.}
\label{fig:signatures-time}
\end{figure}

Apart from the energy signature of the neutrons, the timing signatures are relevant as well since we are using a pulsed beam for interrogation. For neutrons the time-of-flight (TOF) $t_{\rm tof}^n$ is considerable,
\begin{equation}
t_{\rm tof}^n=L/\sqrt{2E_n/m_n}=7.23\times10^{-2} \left(\frac{L}{[\mathrm{m}]}\right)\,\left(\frac{[\mathrm{MeV}]}{E_n}\right)^{1/2}\,\mu\mathrm{s}\,,
\end{equation}
and even the photon time-of-flight is non-negligible, with $t=L/c$
about \SI{3.3}{\micro\second} per km. This has a number of consequences. First,
prompt fission neutrons will be well separated in time from the
backscatter of the interrogating beam, at $2L/c$ or
$\simeq\SI{6.6}{\micro\second}$ for $L=1$\,km, whereas a fission neutron with an
energy of 2\,MeV arrives at 51\,$\mu$s. As we will discuss in detail
in Sec.~\ref{sec:detector} this allows for a clean separation of the
neutron signal despite its much smaller amplitude compared to the beam
backscatter. It also means that firing of the beam itself and any
radiation effects of the LWFA next to the detector are well separated
in time. Further, there is a correlation between the neutron energy and
its TOF that can be used to separate the signal from environmental
backgrounds. The resulting time structure of the neutron signal is
shown in Fig.~\ref{fig:signatures-time} for the three groups of
neutrons defined by how they show up in the detector. The hard
neutrons have energies around 1\,MeV and more and will result in two or
more proton recoils that can be resolved and measured. These measured
recoil energies can be used via kinematics to determine the actual
neutron energy. Then we have soft neutrons, where we cannot
distinguish the various proton recoils and hence cannot kinematically
reconstruct the neutron energy, but still there is a correlation
between the total observed recoil energy and the true neutron
energy. Finally, capture neutrons show only a capture signature, and
we only know that they are neutrons; their energy is around 100\,keV
or less.  The left panel is for W1 whereas the right panel shows a
60\,kg depleted uranium sphere. The hard and soft neutron signals are
very similar in amplitude (by design) and their time structures are
somewhat wider for W1 due to secondary fissions being present. The
capture neutrons show a much longer tail in time for W1, a result of
the moderation from the high-explosive. Overall, the statistics in the
neutron channel is insufficient in most cases to cleanly separate
these two distributions on top of the intrinsic broadening from the
time-of-flight. The gamma signal arises from prompt fission gammas and
fast isomeric fission fragment decays (mostly with lifetimes less than
$100\,$ns) and is spread in time only by the time distribution of the
fission events, since all photons share the same time-of-flight. As we
can see, W1 shows a long-term gamma emission spanning 1\,ms, tracking
the occurrence of fissions over this time period, whereas the DU case
practically stops emitting gammas after \SI{10}{\micro\second}. This provides a
clean signature of fission multiplication and can be exploited for the
unambiguous identification of a close-to-critical mass of fissile
material.

\begin{table}
\caption{Detected counts per pulse of Fig.~\ref{fig:signatures-time}, in total and arriving
after the \SI{1}{\milli\second} shot period, as well as the environment counts in one such period.
}
\label{tab:late}
\begin{ruledtabular}
\begin{tabular}{lccc}
channel & total per pulse & after \SI{1}{\milli\second} & environment per \SI{1}{\milli\second} \\
\colrule
neutrons, hard & \sigHardTot & \num{\sigHardLate} & \sigEnvHard \\
neutrons, soft & \sigSoftTot & \num{\sigSoftLate} & \sigEnvSoft \\
neutrons, capture & \sigCapTot & \sigCapLate & \sigEnvCap \\
gammas & \sigGamTot & \sigGamLate & \sigEnvGam \\
\end{tabular}
\end{ruledtabular}
\end{table}

This figure is drawn out to only 1\,ms, since the beam repetition rate is
never less than 1\,kHz for the configurations we consider. In
Tab.~\ref{tab:late} we show the evolution of the signals beyond the
shot period at 1\,kHz repetition rate and how these numbers compare to
the environmental background. It can be seen quite clearly that only
capture neutrons with a signal/noise ratio of about 0.5 would be able
to accumulate to a statistically significant signal, especially for
higher repetition rates. At a few milliseconds all prompt fission signals
end and only long-lived isomers (rare) and beta-delayed decays (slow)
keep producing a signal. Beta-delayed decays can produce neutrons and
gammas out to several seconds, but at a rate that is totally dwarfed
by the environmental backgrounds.

So far we have assumed that the inspection and target satellites have
zero relative motion, and small relative motion will not affect the
detectable signal much; we should note, however, that a thermal neutron has a
speed of 2.2\,km/s, whereas relative orbital speeds in LEO can reach
14\,km/s. Therefore, we have to account for relative motion for the
case where the relative speed $v_{\rm rel}$ is comparable to or larger
than the neutron speed. All involved speeds are small enough to be
governed by Galilean physics and a detailed calculation can be
found in App.~\ref{app:kinematics}. The key outcome is that slow
neutrons will either take too long to reach the inspection satellite
or get scooped up in a snowplow-like effect. Also, the relative speed
effectively changes the neutron energy: it increases while the range is
decreasing, and vice versa. These effects are fully included in the following.

\begin{figure*}
\includegraphics[width=\textwidth]{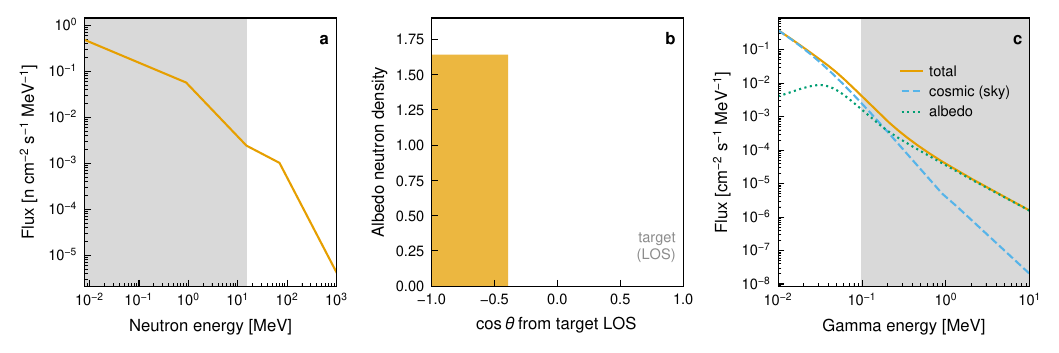}
\caption{The environmental backround. (a) The angle-integrated
atmospheric-albedo neutron flux at a solar modulation of \SI{\solarPhiMV}{\mega\volt}. (b) The albedo neutron density against $\cos\theta$ from the target line of sight. (c) The ambient gamma flux, split into its cosmic
(sky) and albedo terms. Shading marks the span of each channel's detector response.}
\label{fig:backgrounds}
\end{figure*}

The signal consists of neutrons and gammas, therefore we need to
address the environmental background of neutrons and gammas in
LEO. Neutrons are not stable and hence do not occur in cosmic rays but
originate from the interaction of cosmic-ray charged particles,
protons and heavier nuclei, with the Earth's atmosphere, leading to a
neutron albedo. We use the model of Kole {\it et al.}~\cite{kole_env}.
The gamma-ray background has a direct component from galactic x-rays
and an albedo component from the Earth's atmosphere, and we
use the model of Cumani {\it et al.}~\cite{cumani_env}. Both
background components do depend (to differing degrees) on the charged
cosmic-ray flux and hence are sensitive to solar activity and
geomagnetic effects. We do not include the impact of trapped charged
particles and in particular we do not model the South Atlantic
Anomaly. The directionality of the albedo component is accounted for
in the detector response modeling. The energy spectra and angular
dependence of the albedo component are shown in
Fig.~\ref{fig:backgrounds}. We account for the variation of the albedo
components over the orbit by dividing each orbit into 24 equal-exposure
segments, evaluating the detection likelihood for each segment
separately, and summing the result. In determining a detection we use
the lowest-background segments first.

The natural radiation environment in LEO is not only a source of
backgrounds, it also requires satellites to be resistant to the
radiation dose accumulated from exposure to this
environment. Depending on the amount of shielding, average doses for a
\SI{500}{\kilo\metre} altitude, \SI{51}{\degree}
inclination orbit range from 4--20\,Gy per year for aluminum
shielding from 5 down to 2.5\,mm~\cite{nasa_tid}. Solar activity and
solar storms can add many tens of gray in a short period of
time~\cite{Hands:2018,Berger:2018}. Over a five-year period, a typical
mission design value, 20--100\,Gy are accumulated. We therefore assume
that an interrogation dose of 0.25\,Gy over a five-year period is
permissible for the general case, and that where a secondary
inspection is needed 2.5\,Gy is permissible. With $10^9$ electrons per
shot, even at a 100\,kHz repetition rate, the dose rate during
interrogation is small enough not to lead to single-event upsets in
electronics.

\section{Detector and detected signals}
\label{sec:detector}


We need a detector that can be efficient for neutrons from thermal to
MeV energies and to gamma rays from about 0.1--4\,MeV and that can
provide clear particle identification. The detection of reactor
neutrinos poses very similar challenges and recent advances in
detector technology provide 6--7 orders of magnitude of background
suppression \cite{Haghighat:2018mve,MobileAntineutrinoDemonstratorProject:2025sna,LiquidO:2019mxd}. The key
ingredients in these novel detectors are: a high light-yield organic
scintillator, ${}^6$Li as neutron tagging agent, spatial segmentation
and some level of pulse shape discrimination. The three detectors used
as an example use different technological implementations of these
concepts, which for our purposes is not critical. Specifically, we
assume a plastic scintillator loaded with
\SI{\fpeval{round(100*\liFraction,0)}}{\percent} $^6$Li by weight and
segmented into \SI{\voxelCm}{\centi\metre} voxels. The thickness of
the detector is 0.12\,m and we will use detector areas of 1 and
10\,m$^2$, corresponding to detector masses of 120 and 1,200\,kg,
respectively. We assume a light yield of 250 photo-electrons per MeV
of electron-equivalent energy deposition. We model scintillation
quenching with a Birks--Chou model with the correlated coefficient
pair measured on this scintillator~\cite{Awe:2021}; this is
particularly important for proton recoils. We assume a detection
threshold per voxel of 5 photo-electrons, corresponding to
20\,keVee. We further assume that triple pulse shape
discrimination can separate the neutron tag
${}^6\mathrm{Li}+n\rightarrow \alpha + {}^3\mathrm{H}$, proton recoils
and electron/gamma-like energy depositions. We use
GEANT4~\cite{Agostinelli:2003} to compute the true energy depositions in each voxel and
construct detector response matrices relating true to reconstructed quantities.

\begin{figure}
\includegraphics[width=\textwidth]{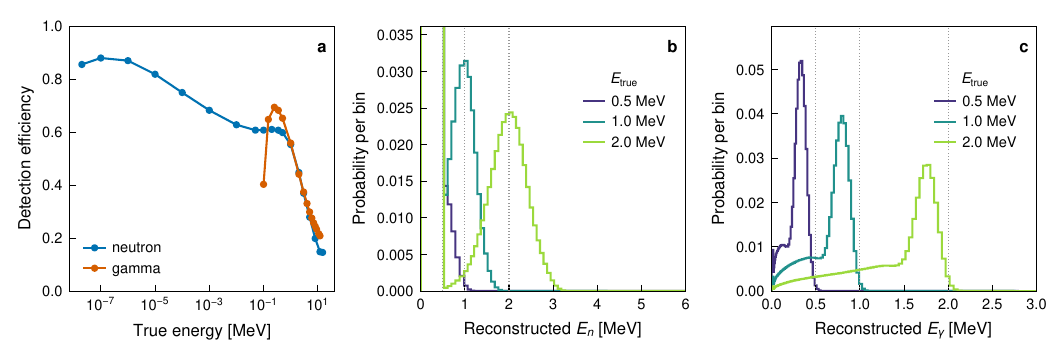}
\caption{The response of the detector.  (a) Neutron and gamma
  detection efficiency against true energy. Panels (b) and (c) show
  the reconstructed energy at three true energies for neutrons and
  gammas, respectively, with the true energies marked by dashed
  lines.}
\label{fig:response}
\end{figure}

For neutrons we distinguish three event classes: hard neutrons, for
which we obtain at least two distinct proton recoil events and can
reconstruct the neutron energy from kinematics; soft neutrons, for
which we can measure the recoil energy but cannot resolve the
kinematics; and, finally, neutrons for which we only have a capture
tag. Thermal neutrons do not propagate more than 1\,cm into the
detector and thus are caught in the surface layer, but even hard and
soft neutrons tend to capture closer to the surface where they enter,
leading to a forward--backward asymmetry in interaction depth. The
resulting detection efficiencies and energy resolutions are shown in
Fig.~\ref{fig:response}.

Charged particle backgrounds from cosmic rays are not simulated, but
the tight spatial and timing coincidence required for genuine neutrons
and the excellent tracking capability of the detector lead to
suppression factors of better than 1:1000.

The scintillation lifetime of EJ-260 is
9.2\,ns~\cite{EJ260} and thus blanking the detector for \SI{\blankShot}{\micro\second}
around each shot and again for \SI{\blankReturn}{\micro\second} around
the return of the backscattered flash ensures that no interference from
these bright events occurs.

\section{The Starlink constellation}
\label{sec:starlink}

\begin{figure}
\centering
\includegraphics[width=0.49\textwidth]{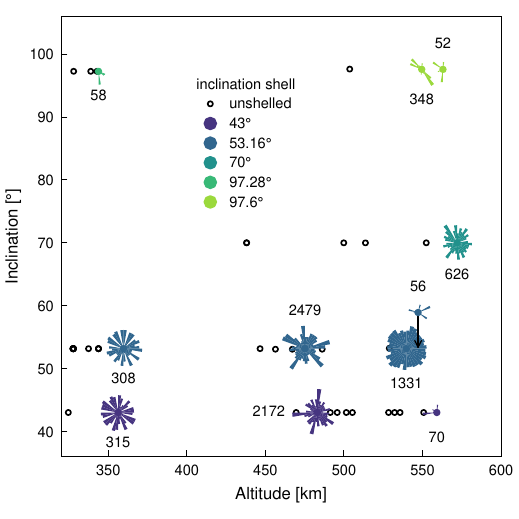}
\caption{The \nSteady{} steady Starlink satellites of the March 25, 2026,
catalog in altitude and inclination.   Each shell is a disk centered on the
point at its core altitude and inclination, with a radius proportional to the logarithm of
the number of its satellites, which is written beside it.  The disk is divided into sectors
of \raanBinDeg$^\circ$ in the right ascension of the ascending node at the catalog epoch,
counted counter-clockwise from the direction of increasing altitude; the length of a sector
is the number of satellites in it relative to the fullest sector of that shell.   The \nUnshelled{}
satellites that belong to no shell are drawn as open circles.}
\label{fig:shells}
\end{figure}

For this work we use Starlink as the reference constellation, since it
is the largest currently in orbit. Specifically, we use two two-line
element (TLE) sets of March 15--16 and March 25--26,
  2026\footnote{\url{https://celestrak.org}}, and we select all
those satellites that have not had a major change in altitude or
inclination, based on the assumption that these are the on-station
satellites. This criterion removed 1{,}832
satellites, which we consider to be in transit either to or from their
station. Overall we retained \nSteady{} satellites, which occur in
five inclination families with \nUnshelled{} satellites not associated
with any of them. We assume that these \nSteady{} satellites perform
station-keeping and that they return to their station after any
collision avoidance maneuver. This real-world constellation differs
from an idealized Walker-type constellation in its greater range of
altitudes and its wide range of inclinations, which as we will see is
relevant for our purposes. The layout of the constellation is shown in
Fig.~\ref{fig:shells} as a function of altitude and inclination.

Thiele {\it et al.}~\cite{Thiele:2026} perform an analysis of natural
close encounters of all cataloged objects in space based on SGP4 with
no maneuvers and no drag (the same assumption we make) and find the
mean time between approaches within 1\,km is 44(36)\,s between all
objects and 66(47)\,s for Starlink satellite to
object. The first number of each pair is from their
direct simulation of the catalog and the number in parentheses from
their analytic rate calculation. Notably, they also report that these
numbers were more than an order of magnitude larger in 2018 before the
deployment of Starlink. They also report
148{,}696 collision avoidance maneuvers for Starlink
between June 1 and November 30 2025, corresponding to about 36 such
maneuvers per satellite per year. Using their numbers and the
assumption that the frequency of encounters scales as the square of the
encounter distance $d$, we infer that the threshold for collision
avoidance is about $d=665-788\,$m. In reality the decision to maneuver
is based on collision risk and not just the miss distance, nonetheless
this indicates that uncontrolled approaches no closer than about
1\,km are routine and do not trigger any concerns.

A typical encounter of the type analyzed above takes place at
kilometer-per-second relative speed and is short, therefore the
relative motion is rectilinear and the relative-position covariance
does not change during the encounter. Following Ref.~\cite{foster1992}
we obtain a collision probability of
\begin{equation}
  P_c \simeq \frac{R^2}{2\sigma^2}\,\exp\!\left(-\frac{d^2}{2\sigma^2}\right)\,,
  \label{eq:pc}
  \end{equation}
where $R$ is the combined hard-body radius and $\sigma$ the position
uncertainty in the plane normal to the relative velocity. Note that
the relative speed does not appear in this equation. For two Starlink
V2 mini this would be 29\,m, set by the span of their solar panels. The
collision risk is therefore set by $\sigma$, the knowledge of the
relative position of the two objects. To manage the close encounters
we consider here we will need on-board position sensing. Bearing below
$0.1\,\mathrm{mrad}$ and range to centimeters have been demonstrated
against cooperative targets~\cite{avgs,storrm}, and sub-meter accuracy
against uncooperative targets has been achieved with
lidar~\cite{nocerino,grsslidar}.  Even with a relative-position
uncertainty of $\sigma=20\,$m the collision probability would be
negligible, $P_c=2\times10^{-22}$ at $d=200\,$m.

Another mode of encounter discussed in the literature on treaty
verification~\cite{Porteous:2025,danagoulian} is co-orbiting, where the
inspector satellite is placed on the same or a very similar orbit and
keeps a close distance for one or many revolutions. The framework for
this analysis is that of relative orbital elements~\cite{damico};
Ref.~\cite{montenbruck2006} shows that holding the relative
eccentricity and inclination vectors of two co-orbiting spacecraft
parallel guarantees a minimum separation whatever the along-track
uncertainty, and that this holds even after a total loss of
control. These techniques have been flight tested on the GRACE
formation~\cite{tapley2004,montenbruck2006} and are in general use for
satellite formation flying.

\section{Orbit mechanics \& Encounters}
\label{sec:orbits}

Satellite motion in the Earth's gravitational field is described to first
order by solutions to the Kepler problem. Angular momentum
conservation keeps each orbit on a stationary plane, perpendicular to
the angular momentum vector. In reality there are many corrections to
this simple picture, for LEO by far the most important is so-called J2
dynamics that arises from the Earth being an oblate spheroid. The
equatorial bulges exert a torque that leads to a precession of the
orbital plane~\cite{vallado}. With inclination $i$ and right ascension of the
ascending node $\Omega$ the unit specific-angular-momentum vector is given by
\begin{equation}
  \hat h = (\sin i\,\sin\Omega,\; -\sin i\,\cos\Omega,\; \cos i)\,.
  \label{eq:hhat}
\end{equation}

The angle between two orbital planes is
\begin{equation}
  \cos\theta = \hat h_1\cdot\hat h_2 = \cos i_1\cos i_2 + \sin i_1\sin i_2\cos(\Omega_1-\Omega_2)\,,
  \label{eq:planeangle}
\end{equation}
and an impulsive rotation of the plane through $\theta$ at a satellite speed $v$ costs
\begin{equation}
  \Delta v = 2v\sin(\theta/2)\;\simeq\; v\,\theta \quad (\theta\ll1)\,,
  \label{eq:planechange}
\end{equation}
about \SI{130}{\metre\per\second} per degree at LEO ($v\simeq\SI{7.5}{\kilo\metre\per\second}$).

The $J_2$-term regresses the node secularly at a rate
\begin{equation}
  \dot\Omega = -\tfrac{3}{2}J_2\left(\frac{R_\oplus}{p}\right)^{2} n\cos i\,,
  \qquad n=\sqrt{\mu/a^3}\,,\quad p=a(1-e^2)\,,
  \label{eq:nodalrate}
\end{equation}
so $\hat h$ precesses about the polar axis at a rate set by $a$ and $i$.  At fixed
$i$ and $e=0$,
\begin{equation}
  \frac{\partial\dot\Omega}{\partial a} = -\frac{7}{2}\frac{\dot\Omega}{a}
  \quad\Longrightarrow\quad \Delta\dot\Omega \simeq -\frac{7}{2}\,\dot\Omega\,\frac{\Delta a}{a}\,.
  \label{eq:nodallever}
\end{equation}
We can use this nodal regression to our advantage, for two satellites at different altitudes the distance between their nodes changes at a constant rate. Thus by changing the altitude, which is a fuel-efficient maneuver, we can close the separation of nodes. The inspector transfers to a \emph{drift orbit} displaced by $\Delta a$, waits there while the node gap closes,
and returns. The cost and the waiting time are
\begin{equation}
  \Delta v_{\rm drift} \simeq n\,\Delta a\,,\qquad
  T_{\rm wait} = \Delta\Omega\,/\,|\Delta\dot\Omega|\,.
  \label{eq:driftorbit}
\end{equation}
The propellant cost is set by $\Delta a$ alone; the node gap sets only
the waiting time.  $J_2$ does not change $i$ secularly, so the
inclination component of $\hat h$ has no such mechanism and must still
be bought at the rate of Eq.~\ref{eq:planechange}. Sun-synchronous targets are the exception: their
inclination is chosen so that $\dot\Omega$ matches the mean motion of
the Earth about the Sun.  That condition fixes the rate rather than
the orbit, so any two sun-synchronous orbits precess identically
whatever their altitudes, $\Delta\dot\Omega=0$, and their node
separation never closes.

From these general principles three different inspection modes with corresponding platforms S1, S2 and S3, can be
designed: S3, where a set of free-falling inspectors is moved by
nodal regression through all orbits within an altitude band;
encounters are fast cross-track fly-bys.  The closest approach is
bounded from below by the altitude spread of the band. S2,
where a set of free-falling inspectors is moved by nodal regression but
the closest approach distance is now set by a specific maneuver and/or
altitudes are adjusted as well; encounters are fast cross-track
fly-bys. S1, where the inspector is brought onto the same orbit
as the target and follows the target at a fixed distance for one or
several orbits. It exploits nodal alignment to minimize the cost of
plane transfer as explained above; encounters happen at near zero
relative velocity. An inspector fleet generally can employ platforms
in each category, but the various encounter types with their
corresponding closest approach distances and dwell times affect the
requirements for the LWFA and the detector as well as inspector design
in terms of fuel and propulsion.

For S3 the whole design is the initial state of each platform: an
altitude and a node, where the altitude is decisive. Two orbits of
different inclination always cross, and at the crossing the separation
is essentially the difference in orbital radius, so a passive platform
produces useful approaches only against targets within a narrow band
of its own altitude. Each platform's altitude offset from its targets
makes its plane precess at a different rate, so relative to every
target in its band it sweeps all node angles within the mission by
itself.  The platforms carry no propulsion and all sit at the
inclination of the densest shell; spreading them in node only keeps
them from presenting the same geometry at the same time.  To size
the fleet we add a platform to whichever altitude cluster is worst
served, re-fly, and read the union coverage, defined by the per-target
best approach over all platforms.
The four S3 platforms of Sec.~\ref{sec:statistics} sit at
\SI{\sThreeInc}{\degree} at \SI{\sThreeAltA}{\kilo\metre}, \SI{\sThreeAltB}{\kilo\metre},
\SI{\sThreeAltC}{\kilo\metre} and \SI{\sThreeAltD}{\kilo\metre}, with nodes of
\SI{\sThreeRaanA}{\degree}, \SI{\sThreeRaanB}{\degree}, \SI{\sThreeRaanC}{\degree} and
\SI{\sThreeRaanD}{\degree} at the epoch of the catalog.

S2 and S1 are routed.  An S2 flies its trajectory forward, screens it for close
approaches to the targets not yet served, and buys a pass with a small burn about a day before
closest approach, which sets the design miss distance; when no target is within reach it changes
altitude to re-tune its precession rate.  An S1 tours a group of targets that share an inclination
and are close in altitude: each leg is a transfer onto a drift orbit, a coast while the
differential precession closes the node gap, and a transfer onto the target's orbit, and the next
target is the one that minimizes propellant plus an exchange rate $\lambda$ times time.  All routes
are flown with Orekit~\cite{orekit} as finite-thrust arcs against the catalog propagated with
SGP4~\cite{vallado2006sgp4}.  The algorithms are described in App.~\ref{app:routing}.
The S2 of Sec.~\ref{sec:statistics} starts at \SI{\sTwoInc}{\degree} with its
node at \SI{\sTwoRaan}{\degree} at the epoch of the catalog, at the altitude of the most
populated \SI{10}{\kilo\metre} band of its targets; over the mission its \sTwoHops{} altitude
hops take it between \SI{\sTwoAltLo}{\kilo\metre} and \SI{\sTwoAltHi}{\kilo\metre}.

Finally, we assume a total mission time frame of 5 years, a total
$\Delta v$ of \SI{5000}{\metre\per\second} and a specific impulse of
$I_{\rm sp}=\SI{1500}{\second}$ comparable to the
  flight-proven SPT-100 family of Hall thrusters, which deliver
  \SIrange{1600}{1700}{\second}~\cite{mitrofanova2011}.  By the
rocket equation this budget costs a propellant load of $e^{\Delta
  v/(I_{\rm sp}g_0)}-1 = \SI{40}{\percent}$ of the
dry mass. It turns out that with those choices time and propellant are
about equally often the binding constraint.


We have two types of encounters, the co-orbit, with a constant
distance and dwell time of minutes and the fly-by, with a quickly
changing distance $\sqrt{v_{\rm rel}^2t^2+b^2}$, where $b$ is the
impact parameter. The median relative speed is
\SI{\surveyVrel}{\kilo\metre\per\second} and thus the typical
encounter (where we fire while the bearing is within
$\firingHalfAngle^\circ$ of closest approach and the distance below
\SI{\lMaxKm}{\kilo\metre}) lasts only 0.1-0.6\,s. Note, that the
total event rate is roughly proportional to $b$ and not $b^2$, since
small $b$ also implies a shorter dwell time. For
the co-orbit a practical bound for dwell time arises from the dose
delivered to the target, to deliver
\SI{\doseCap}{\gray} it takes
\SI{\dwellCap}{\second} for a distance of up to
$L_{\rm fill}=\SI{\lFill}{\kilo\metre}$. Since this
time is rather short compared to the orbit period the parts of the
orbit can be selected that have the smallest background. To evaluate
the range for a possible detection we will use a hypothesis test using the
likelihood ratio as test statistic and choose a detection probability
$P_D=0.9$ and a false positive rate of $\alpha=10^{-3}$, for the
detailed implementation see App.~\ref{app:stats}.

For S1 with a 1\,m$^2$ detector using a 1\,kHz
repetition rate beam the detection range is 1.8\,km.  For S2
satellite with a 10\,m$^2$ detector using a 10\,kHz repetition rate beam
the detection range is 0.7\,km. For S3 with a
10\,m$^2$ detector using a 100\,kHz repetition rate beam the detection
range is 2.9\,km. Note, these ranges match or exceed the avoidance
maneuver threshold derived earlier. The gamma channel used for
secondary screening only, scored on the same shots and dose, reaches
\SI{\gammaReachCoStatic}{\kilo\metre} for S1.

For S1 at fixed dose the number of delivered photons remains
constant when the number of electrons per shot changes and hence, the
dependence on this quantity is weak: the range scales only as the
0.15-th power. For S3 and S2 most of the signal is carried by
out-of-pulse soft and capture neutrons and thus the lever to increase
range is overall more electrons per second, which at fixed charge per
shot means changing the repetition rate: up to 100\,kHz the scaling is
with the 0.6-th power and it flattens quickly after that since the
dead time of up to 3\,$\mu$s from beam blanking starts to bite; at
333\,kHz no detector lifetime remains.

\section{Fleet size and inspection performance}
\label{sec:statistics}

\begin{table}
\caption{\label{tab:platforms} The three inspection platforms, ordered by beam repetition rate.
S1 meets its targets in co-orbit; S2 and S3 in fly-by, S2 routed by its own propulsion and S3
passively at a fixed altitude.  The range is the largest distance at which a single encounter
reaches $P_D=0.9$ at $\alpha=10^{-3}$.}
\begin{ruledtabular}
\begin{tabular}{lcccccc}
platform & repetition rate (kHz) & detector (m$^2$) & $\Delta v$ budget (km\,s$^{-1}$) & encounter mode & range (km) & average power (W) \\
S1 & \fpeval{\repCoorbit/1000} & \fpeval{\faceCoorbit*\faceCoorbit} & \dvCoorbit & co-orbit & \reachCoStatic & \fpeval{round(\dwellCap*\repCoorbit*\electricJ/(\coDaysPerTarget*86400),1)} \\
S2 & \fpeval{\repHzThirteen/1000} & \fpeval{round(\faceSurvey*\faceSurvey,0)} & \dvCoorbit & fly-by, routed & \repRangeThirteen & \fpeval{round(\electricJ*\repHzThirteen*\drFirePerDayT/86400,3)} \\
S3 & \fpeval{\repSurvey/1000} & \fpeval{round(\faceSurvey*\faceSurvey,0)} & \dvSurvey & fly-by, passive & \fpeval{round(\surveyRange,2)} & \fpeval{round(\electricJ*\repHzFourteen*\svFirePerDayT/86400,1)} \\
\end{tabular}
\end{ruledtabular}
\end{table}

We now have all the tools and results to evaluate fleet design, size
and inspection performance: we have a simulation of signatures,
backgrounds and detector performance, and route planning and route
simulation tools to evaluate detection probabilities for all
encounters.  Overall there are a number of choices any verification
regime needs to make. For one, what quantitative threshold would we
like to be able to detect, {\it e.g.} in the IAEA context this is
usually called a significant quantity. In our context this is the
number of deployed warheads. What detection probability do we want to
achieve for this deployed number of warheads? And then finally, over
what time do we want to be able to draw this conclusion? Extreme
choices usually result in either extremely poor overall performance or
exceedingly large resource needs. Also, these choices deeply impact
the technical implementation and may lead to technically
challenging operating points or touch on sensitive national security
issues of the negotiating parties.

The framework presented so far has a robust mechanism to deal with
false positives: they will be referred to secondary inspection by an S1 platform,
which will require a somewhat closer approach,
\SI{\gammaReachCoStatic}{\kilo\metre} at the same dose, or a higher
dose (\SI{\gammaDoseClose}{\gray} at \SI{\gammaReachClose}{\kilo\metre}) to
\emph{unambiguously} verify the absence of fission multiplication. The
false positive rate of our hypothesis test is chosen to be
$\alpha=10^{-3}$ and therefore we will have at most about 10 such
cases. The inspector fleets presented in the following are sized to
absorb this extra workload with ease.

\begin{figure*}
\includegraphics[width=\textwidth]{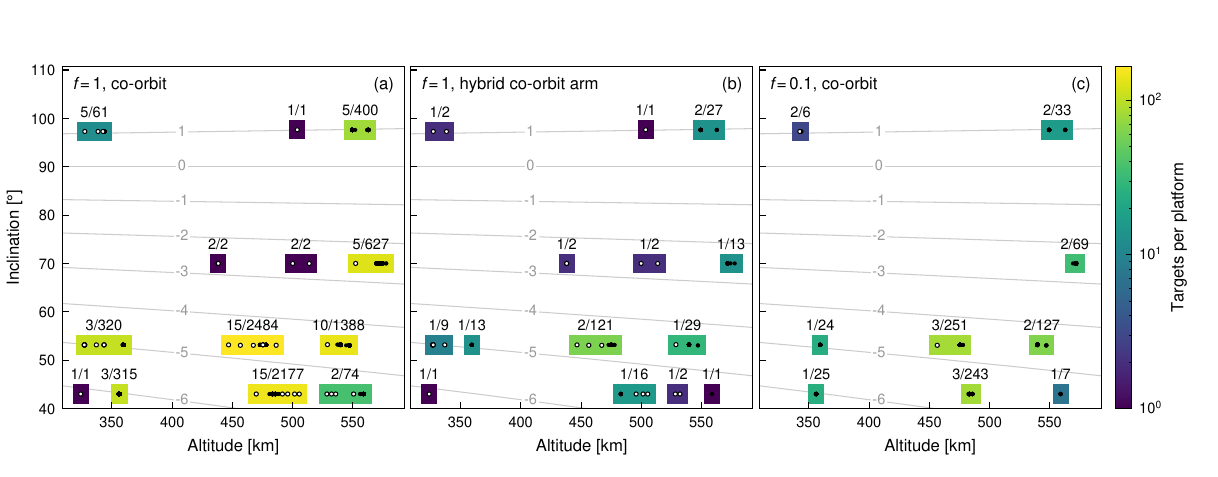}
\caption{\label{fig:sizer-groups} The altitude groups of the fleet sizer for the
three S1 fleets, in the plane of altitude and inclination: (a) the $f=1$ S1 fleet,
(b) the S1 arm of the $f=1$ hybrid and (c) the $f=0.1$ S1 fleet.  Each box is one group, colored
by its targets per platform and labeled platforms/targets; the dots are the targets, open where
the satellite belongs to no shell.  The gray lines are the secular nodal rate in degrees per day.}
\end{figure*}

First, let us review the results for sampling of \emph{all}
satellites. Here the detection probability for a single warhead in the
whole constellation is $P_D=0.9$. We first discuss a fleet of 69
S1 platforms based on a full trajectory simulation. This
S1 fleet serves 10\%, 50\% and 90\% of the targets by day
\fullServeTen, \fullServeFifty{} and \fullServeNinety, respectively
and requires 1821 days, 5 days short of 5 years, to cover all
targets. The efficiency of each of the 69 platforms in terms of targets
served per platform varies widely ranging from 1--192, with a median
of 141. This variation is largely structural, either because the
density of targets on certain orbits is low or because certain orbits
are dynamically more difficult. This is illustrated in
Fig.~\ref{fig:sizer-groups}a, where we show the location in altitude and
inclination of all targets and the groups the routing algorithm
uses; the groups are colored by targets per inspector in
that group. The four worst groups need 1 inspector for each target and
they represent the tail-end of target density. The next worst group
with 5 inspectors for 61 targets is the low-altitude sun-synchronous
group (324\,km, 97$^\circ$). Here the problem is the nodal
regression rate is close to zero but the targets are spread over 6
planes of different RAAN. Changing planes however is prohibitively
slow. Thus this group is inefficient for dynamic reasons. The group at
550\,km shares the same dynamic problem but here the 400 targets sit in twelve
well-populated planes and so no slow plane changes are needed. Verification frameworks
employ a whole toolbox to achieve their task, and this argues for
inspecting targets in these inefficient groups by other means, {\it
  e.g.} on the ground. In this case, referring 67 targets to other
means would save 11 inspection platforms, or in relative terms, 16\%
of effort is expended for 0.8\% of targets.

Full coverage can also be achieved by a hybrid fleet of 4 S3 and 16
S1 platforms, taking exactly 5 years for this task. The 4 S3 platforms serve 10, 50
and 90\% of the targets by days \surveyServeTen, \surveyServeFifty{} and
\surveyServeNinety, and the S1
platforms finish their task in 1583 days. These S3 platforms
require a 100\,kHz repetition rate laser combined with a 10\,m$^2$
detector, but they do not require propulsion beyond station
keeping. Their duty factor is \num{\svDuty} so they would perform the same even
for much larger constellations. The 239 targets or about 3\% not served
by these S3 platforms sit in altitude too far away from any of the
4 S3 platforms. As we can see from Fig.~\ref{fig:sizer-groups}b
the remaining targets follow the same pattern as the full
constellation, just correspondingly less dense. It is much more
economical to serve these targets with S1 platforms. We do
observe again that efficiency ranges widely, from 1 to 67 serves per S1
platform, with a median of 11, compared to 141 for the pure S1
fleet. There are 7 inspectors serving only 1 or 2 targets each, 11 in
total, relative to the 239 targets, we have 43\% of effort for 5\% of
targets. As before, this argues for referring these cases to some
other means of inspection.

In both cases the dose delivered to the target satellites stays below
0.25\,Gy and for S3 the doses remain significantly
smaller than 0.25\,Gy for 99\% of all targets\footnote{There are
\svOverCap{} borderline cases that naively would receive higher doses, but
they can be relegated to S1 inspection within the current fleet
design.}. The peak power requirements are high for S3
with 1,600\,kW but the duty factor is only \num{\svDuty}, i.e. the
average power is only about 4\,W. With moderate propulsion a fly-by
could stay out of the 0.7\,km zone and thus fall within the type of
fly-bys currently experienced by Starlink satellites many times a year
without triggering collision avoidance maneuvers.

Most verification regimes rely on some form of random sampling. Random
sampling from a satellite constellation where we do not re-inspect the
same satellite is statistically the exact equivalent to drawing from
an urn without replacement. Consider a constellation of $N$ satellites
containing $D$ warheads, from which we draw $M\le N$ distinct items
without replacement and inspect each once. A single inspection
recognizes a warhead with probability $\kappa=P_D$ and for simplicity
we assume a zero false positive rate. In practice we achieve that by
secondary inspection, {\it i.e.}, by observing the afterglow in gamma
rays.  We define the sampling fraction $f=M/N$ and for small sampling
fractions $f\ll 1$ the likelihood of a campaign in which no warhead is
found, a ``clean'' campaign, can be approximated as $(1-\kappa
f)^{D}$; the exact expression is derived in App.~\ref{app:sampling}.
Thus even for small sampling fractions we can obtain a reasonable
assurance about the absence of {\it many} warheads. The risk of
discovery $P_\mathrm{dis}(D_1)$ to the violator is simply one minus
the likelihood, that is the probability that for $D_1$ deployed
warheads no clean result is found,
\begin{equation}
\label{eq:risk}
P_\mathrm{dis}(D_1)\simeq\ 1-\left(1-\kappa f\right)^{D_1}\,.
\end{equation}

In the context of a treaty embedding the effect of inspections is not
measured only by the level of assurance of compliance but also by risk
of discovery for a potential violator. Take the following example with
$f=0.1$ and $\kappa=0.9$: while a 43\% assurance\footnote{see App.~\ref{app:sampling} for a derivation.} that no more than 5 warheads
are deployed seems a weak statement, the 37\% probability of discovery faced by a
violator who has deployed 5 warheads may be
fully sufficient to deter. This is of course a well-known
result and the basis for many verification regimes, {\it e.g.}, the IAEA's
system to verify the Treaty on the Non-Proliferation of Nuclear Weapons (NPT)~\cite{ASG}.
The exact posterior exclusion and discovery probabilities for the various sampling fractions
and numbers of warheads are tabulated in App.~\ref{app:sampling},
tables~\ref{tab:posterior} and \ref{tab:discovery}.

To evaluate the fleet size for $f<1$ we randomly draw $f\cdot N$
satellites and task our routing algorithm to determine the number of
S1 platforms to cover this random set. We repeat this exercise
20 times (10 times at $f\geq0.2$), and every target in every draw is
served within the five-year mission. The resulting fleet sizes are listed in the
platforms column of Tab.~\ref{tab:trade} and the scaling clearly is sub-linear for $f\ll1$ for the
same structural reasons discussed.  Fig.~\ref{fig:sizer-groups}c shows
the distribution of inspectors over the orbit groups for $f=0.1$, and
we see that the pattern is nearly unchanged compared to the $f=1$ case
shown in Fig.~\ref{fig:sizer-groups}a.

As previously indicated, the strong disparity in inspector efficiency
is structural and set by orbit dynamics and constellation
density. Assuming that in-space inspection is not the only tool
brought to bear on OST verification we can consider trading space
inspection against some other inspection modality, {\it e.g.} ground
inspection. This trade will be driven in some part by a cost exchange
ratio (CER), how many other inspections can be conducted with the resources
needed for an S1 platform. In reality cost is not measured by
money alone and should be understood in a broader context. We evaluate
how many in-orbit inspectors can be saved for a certain number of other inspections,
dropping the platforms with the fewest targets first; we implicitly assume
that other inspections do not carry the (same) structural cost
disparities as in-orbit inspections do. For all fleet designs the first
platforms are saved for very few other inspections and each further one costs more. In
Tab.~\ref{tab:trade} we show how this trade plays out quantitatively
for a select number of cases at CER of 5, 10 and 20 other
inspections per platform and focus on the size of the space and
other effort. In combination with the choice of sampling fraction a
wide range of fleet sizes, composition and mix with other inspection
modalities appear to be valid approaches to achieve verification of the OST.

Another option at $f=0.1$ is to use active fly-by interrogation. Given
a $\Delta v$ budget of 5,000\,m\,s$^{-1}$ and the same propulsion
parameters as previously, we find that a \emph{single} S2
with only a 10\,kHz beam repetition rate and a 10\,m$^2$
detector can cover 783 of the 785 targets, with the remaining 2
targets covered by a \emph{single} S1 platform, or for
redundancy by a second S2 operated in co-orbit mode, or by other
inspection modalities. Again, this result is based on a full
trajectory simulation. S2 has served 10\%,
  50\% and 90\% of its targets by day \driftServeTen,
  \driftServeFifty{} and \driftServeNinety{} of its 1806-day mission.  The beam duty factor is
below $10^{-7}$, reducing the average power draw to well below 1\,W.

\begin{table}
\caption{\label{tab:trade} The number of in-space S1 platforms and the number of other inspections for various cost exchange rates (CER).  Ranges are over 20 draws (10 at $f=0.2$) for cases where less than the full constellation is sampled ($f<1$).}
\begin{ruledtabular}
\begin{tabular}{lcccccccc}
fleet & targets & platforms & \multicolumn{2}{c}{CER=5} & \multicolumn{2}{c}{CER=10} & \multicolumn{2}{c}{CER=20} \\
 & & & space & other & space & other & space & other \\
\colrule
$f=0.025$ & 196 & 10--14 & 5--7 & 15--39 & 2--3 & 55--108 & 0 & 196 \\
$f=0.05$ & 393 & 13--17 & 9--11 & 5--32 & 5--8 & 44--120 & 2--4 & 153--268 \\
$f=0.1$ & 785 & 16--21 & 13--17 & 0--26 & 10--15 & 18--108 & 7--10 & 119--265 \\
$f=0.2$ & 1,570 & 24--28 & 21--23 & 2--30 & 18--19 & 39--90 & 14--16 & 160--269 \\
\colrule
$f=1$ & 7,852 & 69 & 58 & 39 & 55 & 119 & 51 & 351 \\
hybrid & 239 & 16 & 6 & 46 & 2 & 118 & 0 & 239 \\
\end{tabular}
\end{ruledtabular}
\end{table}

We have demonstrated using accurate route planning and trajectory
physics that a range of inspector fleet architectures can inspect
anywhere from 10--100\% of the real existing Starlink
constellation. Fleet sizes, compositions and platform complexity
vary widely depending on the coverage fraction. Radiation dose with
some moderate level of management can be kept at or below 1\%
(0.25\,Gy) of mission lifetime dose and thus does not represent a real
operational or design constraint. On the other hand, fly-by/co-orbit
distances are in the 0.55--2.9\,km range which is closer than previously
considered~\cite{danagoulian}. The notion of a politically acceptable
distance has no technical foundation; what matters is actual collision
risk and we showed that with on-board position sensors this risk can be
kept negligibly small. This is a key fact that has to enter any
negotiation towards an inspection regime, otherwise the naive
position that only approaches within $X$ kilometers are acceptable
will shift the technical demands for active interrogation towards
infeasibility. The consideration of other inspection modalities that
are not space-based allows a significant reduction of the in-space effort
in recognition of the structural disparities in efficiency of
inspections in space. The key structural factors are constellation
density across the various orbit groups and the slow nodal regression
of sun-synchronous orbits, hence low-density groups and sun-synchronous
orbits are obvious candidates for the other inspection methods.

\section{A minimum treaty framework}
\label{sec:treaty}

A treaty framework is a technical necessity for the concept we
outlined: it would be trivial for a target to evade a prolonged
co-orbit at any distance and for fast close fly-by the usual collision
avoidance framework would make any useful approach impossible. There
are examples of arms-control agreements like START-I (Strategic Arms
Reduction Treaty) or
INF (Intermediate-Range Nuclear Forces)~\cite{START:1991,START:AbA,INF:1987} that came with a complex set of
inspection arrangements, including random sampling of re-entry
vehicles, national technical means, permanent inspector presence at
key facilities and more. There also are treaties that have their own
permanent verification organizations, the NPT~\cite{NPT:1968} has the IAEA
(International Atomic Energy Agency)~\cite{IAEA:Statute} and the
CTBT (Comprehensive Nuclear-Test-Ban Treaty)~\cite{CTBT:1996} has the CTBTO
(Comprehensive Nuclear-Test-Ban Treaty Organization). In particular, the CTBT maintains a
vast technical infrastructure~\cite{CTBTO:PrepCom}, which is a precedent for the in-space
infrastructure our verification concept requires. The specific
organizational framework is outside of the scope for this paper,
nonetheless we would like to outline what provisions any agreement
would need to contain for mere technical reasons. The parties need to agree to the following:

\begin{enumerate}
\item An annual inspection quota and target selection criteria. This includes advance notification of inspection of a given satellite. Depending on the type of orbital maneuvers the lead time can be many days, thus a ``hold-your-orbit'' window pre-inspection is required.
\item A framework to set approach parameters based on the actual collision risk.
\item A real-time protocol to manage approaches including well-defined abort conditions.
\item A declaration of permissible doses and peak dose rates, potentially on a per-target basis.
\item A set of rules when and how to conduct secondary inspections.
\end{enumerate}

As our results indicate, certain orbits like the sun-synchronous
family are hard to inspect in space for basic physics reasons, a small
number of targets requiring a sizable fraction of all effort. Other,
non-space, methods of inspections therefore could reduce the resource
needs for in-space verification considerably. A detailed technical
study of these other inspection methods is outside of the scope of
this work, but it stands to reason that they would likely be a
cost-effective method. At the time the OST was negotiated
on-the-ground inspections were considered problematic because of
classified reconnaissance programs. As a matter of fact,
in the twelve months before the treaty was opened for signature in
1967, \reconLaunches{} of the \orbitalLaunches{} orbital launch
attempts carried a reconnaissance satellite, that is 40\%. Today this
fraction is much smaller, at
\reconPercentNow\%~\cite{McDowell:GCAT}. In any case, on-the-ground
inspection of the inspection platforms likely will be needed to assure
all parties of the capabilities of these systems: positive assurance
will promote compliance and negative assurance will demonstrate that
inspections are harmless. The
treaties cited above all use a varied set of inspection tools and none
relies on a single mode or single technical method. There is no reason
OST verification should deviate from that. 

\section{Summary}
\label{sec:summary}

In this work we made a first attempt at a systems-level approach to
OST verification in space, extending recent work by Porteous~\cite{Porteous:2025}
and Danagoulian~\cite{danagoulian}. We chose high-energy photons of 16\,MeV energy
produced by inverse Compton scattering of a $\sim$1\,GeV electron beam. This
electron beam is to be produced by a laser wakefield accelerator. Both
the accelerator and the scattering photon beam are provided by energy-efficient
diode-pumped fiber lasers delivering \SI{\driveJ}{\joule} per pulse, \SI{\fpeval{\driveJ+\scatterJ}}{\joule} per shot. This technology
has been primarily developed for high-energy physics and has seen dramatic
progress over the past decade~\cite{Kiani:2022}. The laser beam and other
parameters of the required 16\,MeV photon source are close to current
capabilities and the design of a suitable space-qualified fiber laser is
the main R\&D need. Acceleration and photon production happen over
about 10\,cm and the overall system therefore remains compact and
aimable. This technology also results in an overall average electric
power requirement of at most a few watts, and power peaking is no worse than
in modern electric vehicles. We consider three versions of this photon
source with 1\,kHz, 10\,kHz and 100\,kHz repetition rates. These 16\,MeV
photons are well collimated, 0.55\,mrad, and can effectively induce
photofission in actinides while delivering a relatively small dose to
the target. As a result, the beam source cannot damage satellites even
at close range or over prolonged periods of exposure.

This beam source enables a large standoff where the time of flight of
prompt fission neutrons separates them by many microseconds from any
beam interference or backscatter. We anticipate neutron detectors based
on segmented organic scintillators doped with lithium-6 that are derived
from recent advances in neutrino detector technology~\cite{LiquidO:2019mxd,Haghighat:2018mve,MobileAntineutrinoDemonstratorProject:2025sna}. The
detectors provide a high neutron detection efficiency and excellent
charged-particle rejection. We consider detectors of two sizes,
1\,m$^2\times0.12\,$m with a mass of 120\,kg and
10\,m$^2\times0.12\,$m with a mass of 1,200\,kg. We include space
environment neutron and gamma backgrounds~\cite{kole_env,cumani_env} and use GEANT4~\cite{Agostinelli:2003}
to simulate signal generation, detection and background response.

As a sandbox we use the Starlink constellation and select
\nSteady{} of its satellites that are
  in steady operation. We use propulsion parameters:
  a total $\Delta v$ of \SI{5000}{\metre\per\second} and $I_{\rm
    sp}=\SI{1500}{\second}$, for a propellant load of
  \SI{40}{\percent} of the dry mass, that correspond to existing
systems~\cite{mitrofanova2011}. The
two primary encounter types are co-orbit, as in
Refs.~\cite{Porteous:2025,danagoulian}, and
cross-track fly-by with a median $v_{\rm
  rel}=\SI{\surveyVrel}{\kilo\metre\per\second}$. These
different encounter types in combination with propulsion admit
different route planning strategies. A sequential
co-orbit is one in which the inspecting satellite
uses its propulsion to enter the \emph{same} orbit as
the target, stays with the target for one revolution and then moves
onto the next target; this requires matching of orbit
altitude, inclination, node and phase and hence needs
the most $\Delta v$ per
encounter. The routing exploits $J_2$-induced nodal regression to match
the node, trading time for propellant. It further works by
partitioning the constellation into groups that are close in
altitude and inclination and then solves for the
route one satellite at a time resulting in a locally efficient
solution balancing time and fuel. We do not attempt a global solution
of the orbital traveling salesman
problem~\cite{cerf2013,federici2019}, since it is
well known to be NP-hard and a global optimum
presumes knowledge of all satellite positions over
the full mission duration of five years {\it in advance}. The latter
condition almost certainly will not hold for a
realistic constellation. With this routing strategy and the propulsion
parameters chosen, one inspector can serve a median of 141
targets. This is a key result: a single satellite can
inspect a large number of targets \emph{if} they share a close range
in inclination and altitude. These co-orbit encounters can be
engineered to be safe under a total loss of propulsion and on-board
positioning systems like lidar can ensure a
negligible collision probability even for encounter
distances as low as 200\,m. We define inspection
platform S1 to carry a 1\,kHz photon
source and a 1\,m$^2$ detector with a propulsion
system providing a $\Delta v=\SI{5000}{\metre\per\second}$. S1
can perform a co-orbit inspection at a detection probability of
$P_D=0.9$ with a false positive rate of $0.001$ up to standoffs of
\SI{\reachCoStatic}{\kilo\metre} in
\SI{\dwellCap}{\second} with a dose delivered of
0.25\,Gy or less.  A fleet of \fullFleet{} S1 satellites can inspect the entire
constellation of \nSteady{} satellites.

Fly-by encounters are a result of geometry: any two
orbits not sharing the same node and inclination will
cross at a distance determined by their
altitude mismatch. Natural encounters are plentiful
in a dense constellation but usually their phase is not optimal, that
is the two satellites do not meet at the point where the two orbits
are closest and as a result the miss
  distance is much larger than the altitude mismatch. These fly-bys
take place at large relative speeds of up to
14\,km\,s$^{-1}$ and hence provide only sub-second dwell times. In
maneuvered fly-bys the routing algorithm looks ahead to a series of
natural encounters, evaluates the
fuel cost to reduce the miss distance down to the
design value and picks the soonest one that is feasible within the
fuel and time budget. The miss distance can be
engineered to be within 700\,m and again the collision
probability will be negligible
assuming on-board position sensing. We define inspection
platform S2 to carry a 10\,kHz photon source and a
10\,m$^2$ detector with a propulsion system providing a
$\Delta v=\SI{5000}{\metre\per\second}$. S2 can perform a fly-by
inspection at a detection probability of $P_D=0.9$ with a false
positive rate of $0.001$ up to standoffs of
\SI{\repRangeThirteen}{\kilo\metre} in
a pass of \SIrange{0.1}{0.6}{\second} with a dose
delivered of 0.25\,Gy or less. A single S2 platform can inspect
783 of the \tenthTargets{} targets in a random tenth
  of the constellation, the remaining 2 are served by a single S1.

Most verification regimes do not rely on the inspection of every
single item, and also here useful results can be obtained with random
sampling of a fraction of the constellation. For instance, a 10\% random
sample of 785 targets would entail a risk of discovery of 37\% for a
would-be violator that has deployed 5 warheads in space. As just
described, a single S2 platform serving with fly-by encounters can
achieve that and another possibility is to serve these inspections with
17 S1 platforms in co-orbit mode.

As we have pointed out, fly-bys happen naturally as a
result of geometry, and if an inspector has a
powerful enough beam source these natural encounters can be used for
inspection. We define the S3 inspection platform carrying a 100\,kHz
photon source and a 10\,m$^2$ detector without a propulsion system
beyond station keeping. Dividing the constellation into four
altitude bands, we find
that one S3 in each of these bands,
for a total of four platforms, can inspect
all but 239 of the \nSteady{} satellites,
  \SI{97.0}{\percent} of the constellation. These remaining 239
satellites can be inspected by 16 S1 platforms. The coverage gaps
of the four S3 platforms are
structural: they are caused by
satellites sitting at altitudes too far from the
inspection platforms. We also note that the duty
factor of the four S3 platforms is very
small, \num{\svDuty}, so they could
  easily serve future mega-constellations of even
1,000,000 satellites.

Also in the fleet of \fullFleet{} S1
satellites, the number of inspections per
satellite is rather uneven, ranging from as
few as 1 up to 192. This is also
structural, but in this case arises
  either from a low density of targets in a given altitude and
  inclination group or from a nearly zero relative rate of nodal
  regression, as for the sun-synchronous orbits. In sizing and
designing fleets, these structural
issues have to be considered carefully since they
tend to drive the overall fleet size. They also offer the opportunity
for considerable reductions of the in-space effort if other inspection
modalities, like pre-launch inspections on the
  ground, can be implemented. The range of different modalities is
typical for treaty verification and therefore should be
expected for OST verification as well.

All our results for the S1 and S2 platforms, which
have substantial propulsion systems, rely on routing
algorithms we developed for this work. The underlying
problem class is the traveling salesman problem,
notorious for being NP-hard, and thus our solution is
not guaranteed to be optimal and much better
algorithms might be possible. This
is a field of active research, for
instance for satellite servicing~\cite{iskender2026}
or space debris
removal~\cite{cerf2013,federici2019}. However, what
our approach to routing clearly
demonstrates is that the detection and orbit problems
are best treated \emph{together}: {\it e.g.} the
ability to co-orbit many satellites at close distance with one S1
satellite allows us to specify a much less ambitious
laser and detection system. It is this combination
that guarantees useful encounters that can be served with a dose limit
below 1\% of the five-year environmental dose. The
problem is well framed by known orbital mechanics and
the nuclear physics of active interrogation, and thus should be a
fruitful area for the collaboration of experts from both fields.

We also would like to emphasize that an R\&D program towards this
capability could start with a significantly less demanding platform
than S1 in terms of beam parameters. Since it uses co-orbiting
encounters nearly no demands on beam aiming exist and repetition rate can be
traded for dwell time, {\it e.g.} a 100\,Hz beam still would need less
than an hour of beam time at the full range. One could even go further
and work with only $10^8$ electrons per shot at a somewhat reduced range
of about 1\,km. The detector of S1 weighs only 120\,kg and requires
relatively little R\&D to be space compatible. So this S1-light
platform could end up at a size and weight comparable to a Starlink
satellite and even a single S1-light could provide valuable
information in terms of OST compliance. It also would be the clearest
demonstration of the overall feasibility of the concept.

We discussed the minimum treaty requirements: they need to codify the
cooperative nature of approaches for encounters, target selection and
irradiation. Specifically, we need a treaty to specify the annual
inspection quota and target selection criteria. It also needs to set
approach parameters based on the actual collision risk and provide
a real-time protocol for managing approaches. There needs to be agreement
on permissible doses and dose rates. Finally, we need a set of rules
for how to deal with anomalies and secondary inspections.

We demonstrated that comprehensive OST verification
of even very large constellations can be achieved with a limited
number of platforms of moderate size and complexity. We believe this
creates room for the policy community to consider a potential path
towards actual implementation.

\begin{acknowledgments}
I would like to thank A. Danagoulian, C. Geddes, T. Shea, and
K. Walkup for feedback on an earlier version. I would like
to thank the participants of the workshop on Nuclear Problems in the
Atmosphere and Cosmos held on August 20, 2026 at MIT for discussion
and feedback. This work was supported by the U.S. Department of Energy National Nuclear Security
Administration's Enabling Capabilities in Technology (TecH) Consortium under
award number DE-NA0004197. The views and opinions of authors expressed herein
do not necessarily state or reflect those of the United States Government or any
agency thereof.

\end{acknowledgments}

\section*{Data availability}
The orbit data is available at
\url{https://doi.org/10.5281/zenodo.23108548} and interactive viewer
can be found at \url{https://pahubervt.github.io/ost-verification/}.

\section*{Use of AI tools}
AI assistants were used for code development, literature search and
preparation of this manuscript. All physics inputs, simulation choices and
conclusions are the author's own, and every number and reference shown
here was checked against its primary source.

\newpage
\appendix

\section{The ICS photon source}
\label{app:source}

\begin{figure}[h]
\includegraphics[width=0.5\textwidth]{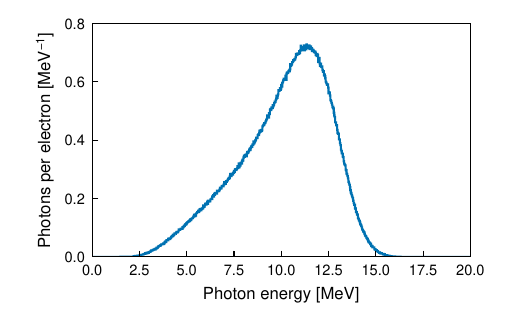}
\caption{The ICS spectrum per electron per MeV inside $\gamma\theta_{\rm lab}\le 1$.}
\label{fig:cone-spectrum}
\end{figure}

Each electron crosses the whole scattering pulse and scatters more
than once. Using Eq.~51 of Ref.~\cite{Rykovanov:2014} we obtain, for a
pulse energy of \SI{\scatterJ}{\joule} focused to a
\SI{\fpeval{round(5*sqrt(\scatterJ/5),1)}}{\micro\metre} spot, a mean
number of scatters $\langle N\rangle = \nScatterMean$. Since there are
many more photons than electrons, the electron beam does not noticeably
deplete the photon beam, and hence the scattering-beam intensity needed
is independent of the charge. For each
scatter we draw one photon at an angle corresponding to the
single-scatter Thomson law, and the electron recoils by exactly that
photon's energy, $\gamma \rightarrow \gamma - E/m_ec^2$, so the later
photons of a history come from a slower electron and are softer.  Over
several scattering events for the same electron the recoil losses are
not negligible: the electron ends at $\langle\gamma_{\rm
  last}\rangle/\gamma_0 = \gammaLastOverZero$ of its initial energy.
For the total yield per electron we obtain $Y_{\rm gen} =
\yieldGenerated$ photons, of which $Y_{\rm cone} = \yieldCone$ lie
inside the cone $\gamma\theta_{\rm lab} \le 1$, a fraction of
$\coneFraction$ of all photons.  Two further effects enter the
Monte Carlo.  The first is the electron pointing, $\sigma_\theta =
\SI{\sigmaTheta}{\milli\radian}$ rms per transverse axis, so that
$\gamma_0\sigma_\theta =
\fpeval{round(\gammaZero*\sigmaTheta/1000,3)}$: the pointing spread is
comparable to the emission cone itself. The second is the energy
spread, $\sigma_\delta = \energySpread$ rms, Gaussian in $\gamma$,
which populates the spectrum above the edge of Eq.~\ref{eq:ics}, as
shown in Fig.~\ref{fig:cone-spectrum}.  We find that the simple
on-axis endpoint of $4\gamma_0^2\hbar\omega_L$ obtained from
Eq.~\ref{eq:ics} is practically identical with the value we find from
the Monte Carlo treatment that accounts for the electron beam energy
and angular spread.  The beam quality enters as yield
and shape: the number of times an electron scatters off the laser
pulse determines the yield, and the focusing of the electron beam
determines the overall opening angle.  In emittance terms,
\SI{\sigmaTheta}{\milli\radian} at a \SI{2}{\micro\metre} spot
corresponds to
\SI{\fpeval{round(\gammaZero*2e-6*\sigmaTheta*1e-3*1e6,1)}}{\milli\metre\milli\radian},
which based on simulations appears feasible~\cite{CapillaryPIC:2025}.

The
drive energy is set by the electron energy: a stage reaches a given energy at
a given intensity and plasma length, and this is to first order independent of the amount of charge that is injected.
As a consequence, the efficiency depends on how close to the optimum charge one can inject, but the pulse intensity itself stays practically unchanged.

It has been shown experimentally that an optimally beam-loaded
accelerator at \SIrange{0.2}{0.3}{\giga\electronvolt} converts
\SIrange{4}{10}{\percent} of the drive pulse into beam
energy~\cite{Gotzfried:2020}; the one directly comparable measurement
at \SI{1}{\giga\electronvolt} extracted about \SI{100}{\pico\coulomb}
from \SI{6}{\joule}, an efficiency of \SI{2}{\percent}, with the
charge limited by injection rather than by
loading~\cite{Grafenstein:2023}; and the scaling discussed in
Ref.~\cite{Lu:2007} puts a \SI{1}{\giga\electronvolt} stage at
\SIrange{6}{13}{\percent}. With the assumptions made here, the beam
carries
\SI{\fpeval{round(\coElectronsShot*\beamEnergy*1.602e-13,2)}}{\joule}
per shot, and for the range of laser-to-beam efficiencies found in the
literature this translates to a range of pulse energies of
\SIrange{0.75}{7.5}{\joule}; we take \SI{\driveJ}{\joule} as a middle ground, noting that this
number could be substantially smaller as well.

\section{Catch-up kinematics}
\label{app:kinematics}

Here we describe how the neutrons propagate between the two satellites at orbital speeds. We do this under the approximation that both satellites represent inertial frames; we solve this problem in the rest frame of the target satellite.  The detector then moves on the straight
line $x(t) = (b, v t)$, so
\[
  R(t) = \sqrt{b^2 + v^2 t^2}, \qquad \dot R(t) = \frac{v^2 t}{R(t)},
\]
and $R$ is minimal ($=b$) at $t = 0$. Here $t$ is absolute encounter time, $b \ge 0$ the impact
parameter and $v \ge 0$ the relative speed; $v = 0$ is the exact static limit.

A neutron of speed $v_n$ emitted isotropically from the origin
at $t_{\rm em}$ is intercepted at the arrival time $t_a$ satisfying $v_n (t_a - t_{\rm em}) =
R(t_a)$; we define the flight time $\tau = t_a - t_{\rm em}$. Squaring
the condition yields
\begin{equation}
  D_2\,\tau^2 - 2 v^2 t_{\rm em}\,\tau - R(t_{\rm em})^2 = 0,
  \qquad D_2 = v_n^2 - v^2 ,
  \label{eq:e21}
\end{equation}
and we define
\begin{equation}
  \Delta = (v^2 t_{\rm em})^2 + D_2 R(t_{\rm em})^2 = v_n^2 R(t_{\rm em})^2 - v^2 b^2 ,
  \label{eq:e22}
\end{equation}
to yield the roots $\tau_\pm = (v^2 t_{\rm em} \pm \sqrt{\Delta})/D_2$. Squaring admits spurious roots; a root is a physical arrival only if it is finite and strictly positive. Care needs to be taken near $D_2\rightarrow0$, i.e.\ $v_n=v$; there are three cases:

\begin{itemize}
  \item $v_n > v$ ($D_2 > 0$): $\Delta \ge b^2(v_n^2 - v^2) > 0$ and the root product
        $-R^2/D_2 < 0$, so exactly one root is positive, the $+$ branch. In the static limit
        $v = 0$ this is $\tau = b/v_n$.
  \item $v_n < v$ ($D_2 < 0$): the root product is positive and the root sum
        $2v^2 t_{\rm em}/D_2$ is positive only for $t_{\rm em} < 0$, so both roots are arrivals
        only for
        \begin{equation}
          t_{\rm em} \le t_{\rm e,max} = -\,\frac{b\sqrt{v^2 - v_n^2}}{v\,v_n} < 0 .
          \label{eq:e24}
        \end{equation}
        The two merge at $t_{\rm em} = t_{\rm e,max}$, where $\Delta = 0$: the integrable caustic.
        Emission later than $t_{\rm e,max}$ never arrives. 
  \item $v_n = v$ ($D_2 = 0$): Eq.~\eqref{eq:e21} degenerates to
        $\tau = -R(t_{\rm em})^2/(2 v^2 t_{\rm em})$, an arrival only for $t_{\rm em} < 0$.
\end{itemize}

The geometric weight, the relative speed
and the incident energy are
\[
  w_{\rm geo} = \frac{A_{\rm det}\,|v_{\rm rel}(t_a)|}{4\pi R(t_a)^2\,|v_n - \dot R(t_a)|},
  \qquad
  |v_{\rm rel}|^2 = v_n^2 + v^2 - 2 v_n v\,\frac{v t_a}{R(t_a)},
  \qquad
  E_{\rm inc} = E\left(\frac{|v_{\rm rel}|}{v_n}\right)^2 .
\]
Here $|v_{\rm rel}|$ is the neutron speed relative to the detector: the neutron travels along
$\hat x(t_a)$ at speed $v_n$, the detector velocity is $(0,v)$, and
$\hat x(t_a)\cdot(0,1) = v t_a / R(t_a)$. The factor $1/|v_n - \dot R|$ is the Jacobian of the
emission-to-arrival map, equal to $|{\rm d}t_a/{\rm d}t_{\rm em}|/v_n$ since differentiating the catch-up condition gives ${\rm d}t_a/{\rm d}t_{\rm em} = v_n/(v_n - \dot R(t_a))$; it
diverges integrably at the caustic, where $\dot R(t_a) = v_n$. In the static limit $v = 0$ it reduces to
$w_{\rm geo} = A_{\rm det}/4\pi R^2$ and $E_{\rm inc} = E$ exactly.

For photons that travel at the speed of light $c$, the relative motion of the two satellites can be neglected since $v\ll c$.

\section{The object and bus models}
\label{app:models}

The bus model is a set of homogeneous boxes with the
total mass and envelope of the V2 Mini as stated in SpaceX's filing with the
FCC~\cite{spacex_fcc_2022_exhibitB}, \SI{800}{\kilo\gram}, a $4.1\times\SI{2.7}{\metre}$ body
and two $12.8\times\SI{4.1}{\metre}$ solar arrays, which the model matches to within
\SI{10}{\percent}.  No public document gives the satellite's internal mass split or its
materials; the model's element inventory agrees with the published composition of a typical
LEO satellite~\cite{schulz2025spacewaste} to within a factor of two for the elements that
matter, carrying about \SI{50}{\percent} more copper and aluminum than that inventory and no
titanium, so it likely overproduces photoneutrons.  Its \SI{40}{\kilo\gram} of argon exceeds
the \SIrange{6}{15}{\kilo\gram} that the publicly documented propulsion and delta-v budget of
the satellite imply~\cite{spacex_gen2_techatt_2020,jankovic2026starlinktle,ecss_e_st_10_04c},
and is kept as a conservative choice, since argon contributes a quarter of the bus's
photoneutron yield.

The neutron and gamma emission of each object is taken from GEANT4~\cite{Agostinelli:2003}
simulations in which photofission and the exclusive $(\gamma,1n)$, $(\gamma,2n)$ and $(\gamma,3n)$
channels are evaluated per isotope from the IAEA Photonuclear Data Library~2019~\cite{iaea_pnd},
the prompt fission neutrons and gammas are sampled with FREYA~\cite{Verbeke:2015}, and the
neutron-induced fission chain and the decay of the fission fragments are transported by GEANT4
itself.

\begin{figure*}
\includegraphics[width=0.75\textwidth]{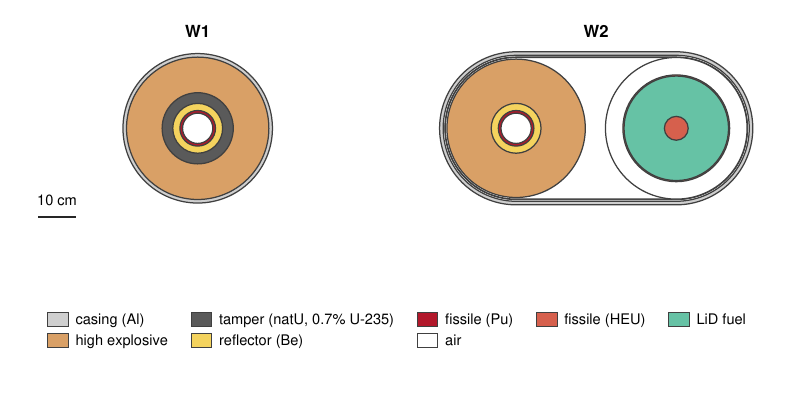}
\caption{Cross-sections of W1 and W2 to scale; details can be found in
  Tab.~\ref{tab:objects}.  Color identifies the material; the foam of
  the W2 secondary is not drawn.}
\label{fig:objects}
\end{figure*}

\begin{table}
\caption{The object models as defined in the GEANT4 geometry, outermost layer first: W1 and
W2 after the notional single-stage and two-stage devices of Sonder et al.~\cite{Sonder:2023},
DU a solid sphere.    Materials in Tab.~\ref{tab:materials}; each mass is the net volume of
the layer times the density of its material.}
\label{tab:objects}
\renewcommand{\arraystretch}{0.8}
\begin{ruledtabular}
\begin{tabular}{llcc}
object & layer & outer radius (cm) & mass (kg) \\
\colrule
W1 & Al casing & \rWOneCasing & \mWOneCasing \\
 & TATB high explosive & \rWOneHe & \mWOneHe \\
 & natural-U tamper & \rWOneTamper & \mWOneTamper \\
 & Be reflector & \rWOneRefl & \mWOneRefl \\
 & Pu pit & \rWOneCore & \mWOneCore \\
 & air cavity & \rWOneCavity & \num{\mWOneCavity} \\
\colrule
W2 & aluminum casing & \rWTwoCase & \mWTwoCase \\
 & air gap & \rWTwoGap & \num{\mWTwoGap} \\
 & natural-U radiation case & \rWTwoLiner & \mWTwoLiner \\
 & cavity & \rWTwoCav & \num{\mWTwoCav} \\
 & primary: TATB high explosive & \rWTwoPriHe & \mWTwoPriHe \\
 & primary: Be reflector & \rWTwoPriRefl & \mWTwoPriRefl \\
 & primary: Pu pit & \rWTwoPriCore & \mWTwoPriCore \\
 & primary: air cavity & \rWTwoPriCavity & \num{\mWTwoPriCavity} \\
 & secondary: foam & \rWTwoSecFoam & \mWTwoSecFoam \\
 & secondary: HEU pusher & \rWTwoSecPusher & \mWTwoSecPusher \\
 & secondary: LiD fuel & \rWTwoSecFuel & \mWTwoSecFuel \\
 & secondary: HEU spark plug & \rWTwoSecPlug & \mWTwoSecPlug \\
\colrule
DU & depleted-U sphere & \rDuBall & \mDuBall \\
\end{tabular}
\end{ruledtabular}
\end{table}

\begin{table}
\caption{The materials of Tab.~\ref{tab:objects} and their densities; isotopic fractions by
weight.  The composition and the density of the materials above the last rule are declared in the
GEANT4 geometry; the three below it are built from GEANT4's NIST material table and
carry its material name.}
\label{tab:materials}
\begin{ruledtabular}
\begin{tabular}{llc p{6.0cm}}
material & used in & density (g\,cm$^{-3}$) & composition \\
\colrule
aged weapons-grade Pu & W1, W2 & \rhoPu & Pu-239/240/241 \isoPuNine/\isoPuZero/\isoPuOne\,\%, Am-241 \isoAm\,\% \\
natural U & W1, W2 & \rhoNatU & U-235 \isoNatU\,\% \\
HEU & W2 & \rhoHEU & U-235 \isoHEU\,\% \\
depleted U & DU & \rhoDU & U-235 \isoDU\,\% \\
TATB & W1, W2 & \rhoTATB & C$_6$H$_6$N$_6$O$_6$ \\
LiD & W2 & \rhoLiD & Li-6 \isoLiSix\,\% of the Li \\
foam & W2 & \rhoFoam & polystyrene, C$_8$H$_8$ \\
\colrule
aluminum & W1, W2 & \rhoAl & G4\_Al \\
beryllium & W1, W2 & \rhoBe & G4\_Be \\
air & W1, W2 & \num{\rhoAir} & G4\_AIR \\
\end{tabular}
\end{ruledtabular}
\end{table}

\begin{table*}
\caption{The bus model: the Starlink V2 Mini~\cite{spacex_fcc_2022_exhibitB} as homogeneous boxes of the
given dimensions, with the element mass
fractions listed.    Total mass \SI{\fpeval{round(\busMassBus+\busMassAvionics+\busMassBattery+\busMassArgon+\busMassAntenna+\busMassWing,0)}}{\kilo\gram}.}
\label{tab:bus}
\begin{ruledtabular}
\begin{tabular}{lcccc p{5.0cm}}
component & box (cm) & placed & density (g\,cm$^{-3}$) & mass (kg) & composition (mass fractions) \\
\colrule
structure & $\busXBus\times\busYBus\times\busZBus$ & \busNBus & \busRhoBus & \busMassBus & Al \busFBusAl \\
avionics & $\busXAvionics\times\busYAvionics\times\busZAvionics$ & \busNAvionics & \busRhoAvionics & \busMassAvionics & Cu \busFAvionicsCu, C \busFAvionicsC, O \busFAvionicsO, Al \busFAvionicsAl, Si \busFAvionicsSi \\
battery & $\busXBattery\times\busYBattery\times\busZBattery$ & \busNBattery & \busRhoBattery & \busMassBattery & O \busFBatteryO, C \busFBatteryC, Al \busFBatteryAl, Cu \busFBatteryCu, Fe \busFBatteryFe, F \busFBatteryF, Li \busFBatteryLi, P \busFBatteryP \\
argon tank & $\busXArgon\times\busYArgon\times\busZArgon$ & \busNArgon & \busRhoArgon & \busMassArgon & Ar \busFArgonAr \\
antenna & $\busXAntenna\times\busYAntenna\times\busZAntenna$ & \busNAntenna & \busRhoAntenna & \busMassAntenna & Cu \busFAntennaCu, C \busFAntennaC, O \busFAntennaO, Si \busFAntennaSi, H \busFAntennaH \\
solar wing & $\busXWing\times\busYWing\times\busZWing$ & \busNWing & \busRhoWing & \busMassWing & O \busFWingO, Si \busFWingSi, Al \busFWingAl, C \busFWingC, H \busFWingH \\
\end{tabular}
\end{ruledtabular}
\end{table*}

\section{Route planning}
\label{app:routing}

An S2 does not need to match a target's plane: two orbits of
different inclination always cross, so a fast pass asks only that the
vehicle and the target reach the crossing at the same radius at the
same moment. What the routing achieves is the timing and the last few
hundred meters of approach.  S2 uses route planning based on a
sequence of flown decisions and the best route is found by iteration.
In each iteration the vehicle's current trajectory is flown forward
over a selection window and screened against every target not yet
served for close approaches inside a capture radius. An approach that
already falls inside the acceptance band is flown
as it is, at no cost. Otherwise the pass is bought with a nudge: a
small burn a day or so before closest approach, solved as a
boundary-value problem onto an aim
point displaced perpendicular to the relative velocity on the
high-altitude side. This ensures that the encounter happens at the design miss
distance. Candidates are solved for in
order of time to closest approach and the soonest affordable one is
taken. Choosing the soonest rather than the cheapest is beneficial, because
the mission time is a more binding constraint than fuel, {\it i.e.} we
usually run out of time before running out of fuel.  A candidate is
refused if the burn would need more thrust time than its assigned
lead time. Once no target within reach can be served cheaply the
alternative is a hop, a low-thrust change of altitude that re-tunes
the precession rate; staying and hopping are compared at each
iteration by evaluating the cost of both. The vehicle retires when
hops stop producing passes.

For S1 we perform route planning inside a group of targets that
share the same inclination and are close in altitude, since here the
inspector must match the target's plane.  A tour never leaves its group,
because inclination cannot be changed cheaply and only within a group
is the node lever usable. Inside a group
the route is built one leg at a time. A leg is a thrust arc onto a
drift orbit, a coast during which the differential precession closes
the node gap, and a second arc onto the target's orbit; its unknowns
are the drift orbit's semi-major axis and inclination and the coast
time, fixed by two closure conditions: one on the node, one on the
phase along the orbit.  Each admits any whole number of extra
relative revolutions, so the leg is solved for every winding pair and
the cheapest solution kept. The next target is chosen from those
nearest in node by minimizing propellant plus a declared exchange rate
$\lambda$ times time; that exchange rate is a cost function that
weighs a day of campaign against propellant in meters per
second. The final approach is solved separately: a minimum-propellant
low-thrust program in relative orbital elements, constrained to stay
outside a keep-out radius, ending in a hold ahead of or behind the
target whose passive safety is measured by propagating both bodies
unthrusted for several revolutions after arrival. The fleet follows
from the route: one inspector flies its group until the mission-time
or propellant cap stops it, the next starts on what is left, and so
forth.  Federici {\it et al.}~\cite{federici2019} pose active debris
removal in exactly this form, for a set of targets sharing an orbital
plane at slightly different altitudes, corresponding to our group, and
solve the sequence and the rendezvous epochs together by simulated
annealing. They impose a tight mission time, which is why they buy
each transfer with impulses rather than waiting, and they note that
the alternative of letting $J_2$ align the planes becomes impractical
under such a constraint.  We have the opposite budget: a five-year
campaign with no date attached to any single target, so the waiting is
affordable and the exchange rate $\lambda$ quantifies this trade-off.

For the sake of numerical efficiency the S1 route planner works
on approximations.  The tour planner represents every orbit by its
mean elements and advances them with the first-order secular rates of
Eq.~\ref{eq:nodalrate}. For the thrust arc we use Edelbaum's
closed-form circle-to-circle low-thrust solution~\cite{edelbaum1961},
with the angles accrued along the arc by quadrature.  Within that
model the two closure conditions are linear in the drift orbit's
offsets in semi-major axis and inclination, so each pair of revolution
numbers is solved in closed form. The small residual left by the
linearization is removed by two or three damped Newton--Raphson
iterations on the exact arcs, and the coast time is left as a
one-dimensional minimization.  The final approach is planned in
relative orbital elements, with the closed-form $J_2$ state-transition
matrix of Chernick and D'Amico~\cite{chernick2018} and a
minimum-propellant linear program over thrust slots.

For numerical integration of all orbits and maneuvers we use
Orekit~\cite{orekit} with an eighth-order Dormand--Prince scheme, the
geopotential evaluated to degree and order four, and every burn as a
finite-thrust arc.  The targets are propagated on the same clock from
the frozen two-line catalog with SGP4~\cite{vallado2006sgp4}, and
the encounters, miss distances and relative speeds quoted here are
measured on those integrated states. We have cross-validated a number
of selected trajectories including maneuvers against
GMAT~\cite{gmat2014}.

\section{Detection statistics}
\label{app:stats}

We define a log-likelihood for Poissonian distributed counts over
time and energy in the detector, specifically we
use for hard neutrons $34\times9$ (time, energy), soft
$34\times19$ (time, photoelectrons), and capture $34$ (time). For
observed counts $n_i$ in cell $i$ with expectation $\mu_{Hi}$ under
hypothesis $H$, the likelihood is
\begin{equation}
  \mathcal{L}_H=\prod_{s}\prod_{i}\frac{\mu_{Hi}^{\,n_i}\,e^{-\mu_{Hi}}}{n_i!},
  \qquad
  \mu_{Hi}(s)=n_{s}\Big[\frac{A_{\rm det}}{L^2}S^{H}_i+b_i(s)\Big],
\label{eq:coo-l}
\end{equation}
the test statistic is the log-likelihood ratio
\begin{equation}
  q=\ln\frac{\mathcal{L}_1}{\mathcal{L}_0}
   =\sum_{s}\sum_{i}\Big[n_i\ln\frac{\mu_{1i}}{\mu_{0i}}-(\mu_{1i}-\mu_{0i})\Big],
\label{eq:coo-q}
\end{equation}
with $n_s$ the shots spent in orbit segment $s$ of 24, $b_i(s)$ the
environment per shot period at that segment's magnetic latitude, and
$S^{H}_i$ the escaping neutrons per shot folded through the detector
response. The likelihood carries only bins
that do not occur when the detector is blanked (2\,$\mu$s after the beam pulse
and 1\,$\mu$s after arrival of the gamma backscatter).

At repetition rate $f_{\rm rep}$ the detector records one periodic
histogram: a neutron arriving a time $t$ after its own gamma
backscatter flash is recorded at the time $t \bmod T$, $T=1/f_{\rm
  rep}$, so the slow tail of one shot arrives during the periods of
later shots.  The time axis of every cell is that time, $T$ divided
into logarithmic bins,
\SI{\fpeval{round(1e6/\repCoorbit,0)}}{\micro\second} into 34 for the
co-orbit and \SI{\fpeval{round(1e6/\repSurvey,0)}}{\micro\second} into
10 for the fly-by.  The fold sums each shot's arrival distribution
over the wraps $k=0,1,2,\ldots$ in closed form.  In a sequence of $n$
shots the events that arrive $k$ periods late are inside the sequence
for $n-k$ periods, so wrap $k$ enters with a weight $\max(n-k,0)$.
The environment is counted for one full period each shot. In the
fly-by the same fold runs per epoch with that epoch's shots, and what
arrives later by more than one period after the firing window closes
is not scored.

The gamma channel is scored by the same construction, with the
following differences. A gamma cell is measured photon energy times
time in the gamma window, $45\times36$ cells for the co-orbit and
$47\times12$ for the fly-by.  The energy axis covers
\SIrange{\gammaWinLo}{\gammaWinHi}{MeV} in bins matched to the
detector's energy resolution; deposits outside the window are not
scored. TOF is trivial for photons and no effects from relative motion
arise.

We perform a hypothesis test using the likelihood ratio (or difference
of log likelihoods) as test statistic. The null hypothesis $H_0$ is
given by the bus only signal, whereas the alternative hypothesis $H_1$
is given by the combined signal of the bus and warhead. Since the
log-likelihood is the sum of many random variables $\mu_{1i}-\mu_{0i}$
the resulting distribution of the test statistic will be normal
distributed (due to the central limit theorem) with mean $-\Lambda$
under $H_0$, mean $+\Lambda$ under $H_1$ and variance $2\Lambda$ under
both; we verified that this approximation is valid with Monte Carlo simulation.  A false-alarm
probability $\alpha$ and a detection probability $P_D$ then fix the
$\Lambda$ required,
\begin{equation}
  \Lambda^\star=\tfrac12\big(z_{1-\alpha}+z_{P_D}\big)^2,
\label{eq:lambda-star}
\end{equation}
with $z_p$ the standard normal quantile.  With $\alpha=10^{-3}$ and $P_D=0.9$,
$\Lambda^\star=\lambdaStar$.


In a fly-by the inspector passes the target on a straight relative
trajectory at separation $R(t)=\sqrt{b^2+v_{\rm rel}^2t^2}$, with $b$
the closest approach, and fires while the bearing is within
$\firingHalfAngle^\circ$ of closest approach and
$R\le\SI{\lMaxKm}{\kilo\metre}$.  Eqs.~\ref{eq:coo-l} and
\ref{eq:coo-q} apply with the following changes.  The sum over orbit
segments $s$ becomes a sum over epochs $t_k$ of the pass inside the
firing window, each with its own shots $n_k$ at the repetition rate
of 100\,kHz and its own distance $R(t_k)$. The $A/L^2$ is
replaced with the full neutron propagation as derived in
App.~\ref{app:kinematics}.  Each shot period is divided into 10 arrival-time bins, and we
use the following energy bins hard, $9$, soft $19$ and capture
$10$. Note that we score shots from highest to lowest likelihood and
would stop the beam once the summed log likelihood over these shots
has reached $\Lambda^*$, this serves to deliver the minimum
possible dose at the minimum possible energy expenditure.

\section{Sampling without replacement}
\label{app:sampling}

We proceed by defining
\begin{equation}
f = \frac{M}{N}\ \ \text{(sampling fraction)},
\qquad
\bb = 1-\kappa f\ \ \text{(escape probability per warhead)}\,.
\end{equation}
``Clean'' denotes a campaign in which no warheads were found. A given warhead survives
the campaign undetected if it is either not drawn, or is drawn and missed, which has marginal
probability $\bb$. For $f\ll1$ the draw is well approximated by one with replacement, the
warheads survive independently, and the likelihood of a clean campaign is
\begin{equation}
L(D) = P(\text{clean}\mid D) \simeq \bb^{\,D}\,;
\label{eq:lik-approx}
\end{equation}
the exact expression for the draw without replacement is Eq.~\ref{eq:lik}.
In this limit everything below keys on the single number
$\kappa f$, the ``effective coverage'': recognition and sampling fraction are
interchangeable, and only their product matters.

With a uniform prior in $d$, the number of warheads deployed, the posterior is geometric with
ratio $\bb$,
\begin{equation}
P(D=d\mid\text{clean}) = \kappa f\,(1-\kappa f)^{d},
\qquad
P(D\leq d\mid\text{clean}) = 1-(1-\kappa f)^{\,d+1}\,.
\label{eq:post}
\end{equation}

With $N$, $D$, $M$ and $\kappa$ as in Sec.~\ref{sec:statistics} and $\bb$ as above, the $D$ warheads are
not statistically independent, because drawing without replacement correlates their inclusion.
Conditioning on the number of warheads that fall in the sample gives the exact likelihood of a
clean campaign,
\begin{equation}
L(D) = P(\text{clean}\mid D)
= \binom{N}{D}^{\!-1}\sum_{j}\binom{M}{j}\binom{N-M}{D-j}(1-\kappa)^{j}
\;\simeq\; \bb^{\,D}\,,
\label{eq:lik}
\end{equation}
with $j$ running over $\max(0,D-N+M)\le j\le\min(D,M)$.
The approximation is the $f\ll1$ limit, Eq.~\ref{eq:lik-approx}. With a uniform prior in $d$, the number of warheads deployed, the Bayesian posterior is
\begin{equation}
P(D=d\mid\text{clean}) = \frac{\kappa\,(M+1)\,d!\,(N-d)!}
{\left[1-(1-\kappa)^{M+1}\right](N+1)!}
\sum_{j}\binom{M}{j}\binom{N-M}{d-j}(1-\kappa)^{j}\,,
\label{eq:exact}
\end{equation}
$j$ ranging as in Eq.~\ref{eq:lik}; for $f\ll1$ it reduces to the geometric posterior of
Eq.~\ref{eq:post}.  The exact posterior exclusion probability is tabulated in
Tab.~\ref{tab:posterior} and the exact discovery probability, one minus Eq.~\ref{eq:lik},
in Tab.~\ref{tab:discovery}.

\begin{table}[h]
\caption{\label{tab:posterior} Posterior probability $P(D\le d\mid\text{clean})$ that at most $d$
warheads are deployed, given a campaign in which none were found, as a function of the
sampling fraction $f$ and of $d$.}
\begin{ruledtabular}
\begin{tabular}{cccccc}
warheads $d$  & at most 1 & at most 5 & at most 10 & at most 20 & at most 50 \\
\colrule
$f= 0.025$ & 0.045 & 0.128 & 0.223 & 0.382 & 0.690 \\
$f= 0.05$  & 0.088 & 0.242 & 0.398 & 0.621 & 0.906 \\
$f= 0.1$   & 0.172 & 0.433 & 0.646 & 0.863 & 0.992 \\
$f= 0.2$   & 0.328 & 0.696 & 0.888 & 0.985 & $>0.999$ \\
\end{tabular}
\end{ruledtabular}
\end{table}

\begin{table}[h]
\caption{\label{tab:discovery} Probability of discovery {\it i.e.}
the probability that a violator who has deployed $D_1$ warheads is found out, as a function of the sampling fraction $f$ and of $D_1$.}
\begin{ruledtabular}
\begin{tabular}{cccccc}
warheads $D_1$  & 1  & 5  & 10  & 20  & 50  \\
\colrule
$f= 0.025$ & 0.023 & 0.107 & 0.204 & 0.366 & 0.680 \\
$f= 0.05$  & 0.045 & 0.206 & 0.369 & 0.603 & 0.901 \\
$f= 0.1$   & 0.090 & 0.376 & 0.611 & 0.849 & 0.991 \\
$f= 0.2$   & 0.180 & 0.629 & 0.863 & 0.981 & $>0.999$ \\
\end{tabular}
\end{ruledtabular}
\end{table}

\newpage
\bibliography{paper,local}

\end{document}